\documentclass[astrosymb,trackchanges]{aastex631}

\usepackage{amsmath}
\usepackage{esint}
\usepackage{multirow}

\shorttitle{Interior heating from stellar flares}
\shortauthors{Grayver et al.}

\begin{document}

\title{Interior heating of rocky exoplanets from stellar flares with application to TRAPPIST-1}

\correspondingauthor{Alexander Grayver}
\email{agrayver@uni-koeln.de}

\author[0000-0003-1132-3705]{Alexander Grayver}
\affiliation{Institute of Geophysics and Meteorology, University of Cologne,\\ 
Albertus-Magnus-Platz,\\
50923 Cologne, Germany}
\affiliation{Institute of Geophysics, ETH Zurich \\
Sonneggstrasse 5, NO \\
8092 Zurich, Switzerland}

\author[0000-0002-0673-4860]{Dan J. Bower}
\affiliation{Center for Space and Habitability, University of Bern \\
Gesellschaftsstrasse 6 \\
3012 Bern, Switzerland}

\author[0000-0003-1413-1231]{Joachim Saur}
\affiliation{Institute of Geophysics and Meteorology, University of Cologne,\\ 
Albertus-Magnus-Platz,\\
50923 Cologne, Germany}

\author[0000-0001-6110-4610]{Caroline Dorn}
\affiliation{Institute of Computational Sciences,
University of Zurich \\
Winterthurerstrasse 19 \\
8057 Zurich, Switzerland}

\author[0000-0003-2528-3409]{Brett M.~Morris}
\affiliation{Center for Space and Habitability, University of Bern \\
Gesellschaftsstrasse 6 \\
3012 Bern, Switzerland}


\begin{abstract}

Many stars of different spectral types with planets in the habitable zone are known to emit flares. Until now, studies that address the long-term impact of stellar flares and associated Coronal Mass Ejections (CMEs) assumed that the planet's interior remains unaffected by interplanetary CMEs, only considering the effect of plasma/UV interactions on the atmosphere of planets. Here, we show that the magnetic flux carried by flare-associated CMEs results in planetary interior heating by ohmic dissipation and leads to a variety of interior--exterior interactions. We construct a physical model to study this effect and apply it to the TRAPPIST-1 star whose flaring activity has been constrained by Kepler observations. Our model is posed in a stochastic manner to account for uncertainty and variability in input parameters. Particularly for the innermost planets, our results suggest that the heat dissipated in the silicate mantle is both of sufficient magnitude and longevity to drive geological processes and hence facilitate volcanism and outgassing of the TRAPPIST-1 planets. Furthermore, our model predicts that Joule heating can further be enhanced for planets with an intrinsic magnetic field compared to those without. The associated volcanism and outgassing may continuously replenish the atmosphere and thereby mitigate the erosion of the atmosphere caused by the direct impact of flares and CMEs. To maintain consistency of atmospheric and geophysical models, the impact of stellar flares and CMEs on atmospheres of close-in exoplanetary systems needs to be studied in conjunction with the effect on planetary interiors.

\end{abstract}
 


\section{Introduction} 
\label{sec:intro}

\begin{figure}[ht!]
\centering
\includegraphics[width=\textwidth]{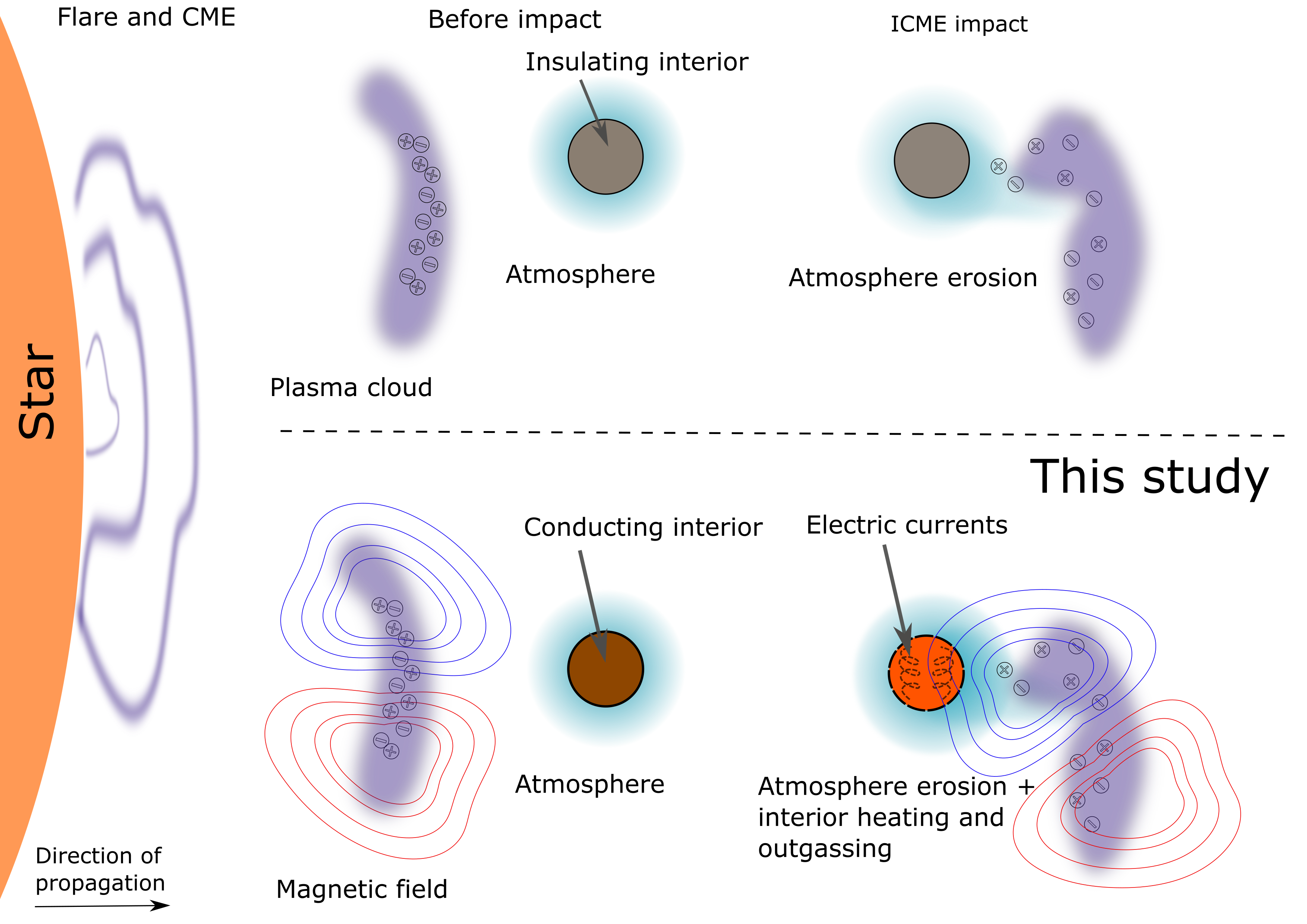}
\caption{The effect of an ICME impact on a planet. Top row shows an ICME interacting only with the exosphere (considered by previous studies), in which the interior is treated as an insulator.  Bottom row shows that planetary interiors are conductive and therefore the magnetic energy carried by an ICME induces currents in the interior, resulting in ohmic dissipation and heating.  Influence of an intrinsic magnetosphere around a planet is not depicted here, but is discussed in the main text. \label{fig:sketch}}
\end{figure}

Stellar flares are among the few observables \citep{paudel2018k2} that allow us to look into the dynamics and evolution of planetary systems and hosting stars. The majority of energetic flares from the Sun are associated with Coronal Mass Ejections (CMEs) \citep{youssef2012relation}, and recent modeling and observations suggest the same holds for other stars \citep{herbst2019solar, moschou2019stellar, argiroffi2019stellar, odert2020stellar}. Therefore, flaring activity of a star is a key consideration for evaluating planetary habitability because plasma bursts and EUV associated with CMEs and flares can lead to ionization of exospheres and facilitate atmospheric erosion \citep{khodachenko2007coronal, airapetian2020impact}.  However, while interactions of flares and flare-associated CMEs with planetary exospheres have been studied extensively, CMEs that impinge on planets will also induce electric currents in the interior and dissipate part of the electromagnetic (EM) energy as heat (Figure \ref{fig:sketch}). modeling and quantifying this heating effect has so far been overlooked in the literature. Presentation of this new heating mechanism is relevant for ongoing searches of atmospheres around rocky exoplanets, including an up-coming dedicated JWST campaign for TRAPPIST-1c \citep{2021jwst.prop.2304K}.

This study presents a physical model that evaluates the amount of heat produced within a planet due to magnetic field variations caused by Interplanetary Coronal Mass Ejections (ICMEs). The model calculates the amount of heat produced over an arbitrary period of time, enabling us to assess the long-term implication of this heating mechanism for the planetary interior and atmospheres. For rocky planets, the underlying physics is governed by Maxwell's equations that fully determine the distribution of electromagnetic field in the planetary interior induced by an arbitrary external electric current. By selecting plausible conductivity models for the planetary interior, we solve the governing equations for an ensemble of ICMEs sampled from a flare frequency distribution, assuming that flare events lead to CMEs. To account for variability and uncertainty in the input variables and parameters, the physical model is posed in a stochastic form, whereby major input and output variables are represented by statistical distributions and thus encompass many different settings and scenarios. This provides a comprehensive picture about the impact of this heating mechanism on rocky exoplanets. 

Additionally, we do not limit ourselves to super-flare events or a class of super-flare stars. Although individual extreme energy flares pose a significant threat to life and can deprive a planet of its atmosphere \citep{tilley2019modeling, airapetian2020impact}, their occurrence rate is low and often accompanied by a large observational uncertainty. Therefore, we avoid drawing the long-term consequences for planetary evolution based solely on rare extreme events. To obtain a statistically more relevant picture, the model developed here relies on the entire observed flare frequency distribution, presently available for many stars through extended observational campaigns \citep{maehara2015statistical, paudel2018k2, Seli2021, ilin2021giant}. Consequently, the results are not biased by considering only extreme events. On the contrary, we show that less energetic events contribute substantially over a long ($\approx 10^5$ years) time owing to their high frequency of occurrence. Note that several physical mechanisms that can induce flare-associated CMEs are known, including processes within active regions or star-planet interactions \citep{shibata2013can}. The heating effect described in this work is, in principle, agnostic to the origin of a CME as long as that the latter propagates the magnetic flux to a planet. However, some of our model parameters are adapted from the solar system where the most energetic flare/CME events originate from active regions \citep{youssef2012relation}.

Another aspect addressed in this work is the role of an intrinsic magnetic field. Sustainable planetary dynamo fields have been proposed to shield planetary atmospheres from erosion by deflecting the incoming ionized plasma, although recent studies show that an intrinsic magnetic field is not the only factor to consider \citep{ramstad2021intrinsic}. While the exact role of an intrinsic magnetic field in sheltering atmospheres from loss remains debated, in this work we show that an intrinsic magnetic field is likely to promote larger interior heating upon interaction of an intrinsic magnetosphere with ICMEs, compared to no magnetic field. To understand these phenomena, we recall that the interplay between stellar-wind and planetary magnetic fields leads to the formation of magnetospheres around planets \citep{baumjohann1996basic}. Intrinsic magnetospheres are complex nonlinear systems and their structure is controlled by many properties, including the geometry and strength of an intrinsic planetary magnetic field and stellar wind properties. When an ICME impacts a planet with an intrinsic magnetosphere field, energy propagates in the magnetosphere leading to the intensification and generation of magnetospheric currents \citep{akasofu1981energy, baumjohann1996basic}. This gives rise to magnetic storms---relatively short periods of large spatiotemporal magnetic field perturbations that impinge on a planet's surface.  Intensification of magnetosphere currents results in energy exchange with the planetary interior via EM coupling, and during periods of large external perturbations the induced subsurface currents have higher density and thus produce more heat.

To first order, the amplification effect of an intrinsic magnetosphere can be quantified by using the relation between interplanetary magnetic field (IMF) and the Disturbance Storm Time (Dst) index for Earth \citep{gopalswamy2008solar, richardson2010near}. The $Dst$ index is a proxy for the strength of the global magnetospheric ring current, describing the horizontal magnetic field induced by an axially symmetric part of the ring current as observed on the Earth's surface \citep{akasofu1981energy}. Since the ring current produces a magnetic field that is opposite to the Earth's main magnetic field, $Dst$ is negative during global magnetic storms. Figure \ref{fig:imf_vs_dst} shows that during magnetic storms, perturbations on Earth's surface are a factor of 3--9 times larger than the magnitude of the interplanetary magnetic field (IMF). This level of amplification is a lower bound since $Dst$ index does not characterize other current-generating regions such as those in the ionosphere or at polar latitudes, which can lead to much stronger magnetic substorms, albeit at smaller scales \citep{finlay2017challenges}. If a rocky planet possesses a dynamo magnetic field of similar magnitude and dipolar structure to Earth, one can adopt the first-order approximate relation shown in Figure \ref{fig:imf_vs_dst} to estimate the peak magnetic field during an ICME impact. 

There exist other types of electromagnetic interactions. \cite{kislyakova2017magma} (see also \citep{bromley2019ohmic}) showed that the motion of a planet through a stellar magnetic field can result in significant interior heating, subject to the geometry of the stellar magnetic field and the star-planet orbital configuration. Although this model also invokes ohmic dissipation on a fundamental level, the driving forces behind their mechanism and our study are distinct. While the source for EM induction heating in \cite{kislyakova2017magma} is a single harmonic due to motion of a planet through an inclined stellar magnetic field, we instead focus on transient excitations due to ICMEs. Specifically, factors that determine the heating rate in their model, such as the stellar magnetic field, synodic period and orbital inclination, have no direct control on the heating rate in our model due to flares (although a strong stellar field can facilitate confinement of CMEs as we discuss later).
Other types of EM interactions, for instance due to a direct electrodynamic coupling of close-in planets with the star \citep{laine2011interaction, fischer2019}, also potentially lead to induction heating, although the amount of heat produced in these scenarios has not been studied.  In short, interior heating due to ICMEs is a distinct mechanism worthy of investigation, particularly given current and future observational campaigns that will constrain the flaring activity of nearby stars and potentially detect the CME events directly.

\section{Data and Methods}
\label{sec:methods}

We apply our model to the M8V star TRAPPIST-1, the host to seven Earth-sized transiting planets, of which three are expected to reside in the habitable zone (HZ) \citep{agol2021refining}. Despite the relatively old age of TRAPPIST-1 near 7 Gyr \citep{Burgasser2017}, it is a magnetically active flare star \citep{paudel2018k2}, as is common among the latest-type stars \citep{Reiners2010,Seli2021}. The magnetic activity of TRAPPIST-1 is poorly understood, as its rotational modulation and absolute spectroscopy appear to be influenced by bright regions \citep{Morris2018}, but photometry and transmission spectroscopy have yet to reveal evidence for contamination by stellar heterogeneity \citep{Morris2018b, Garcia2022}. No matter how its surface magnetic activity is expressed, its flares are ubiquitous in photometry from the optical \citep{Vida2017,ducrot2020trappist} to the infrared \citep{Davenport_2017}.

The orbital distances and radii of the TRAPPIST-1 planets were obtained from \cite{agol2021refining}. Parameters of the flare frequency distribution for TRAPPIST-1 were taken from \cite{paudel2018k2}, where TRAPPIST-1 was observed by Kepler for 70.6 days, detecting 39 flare events with energies between $0.21\cdot 10^{30}$ erg and $230\cdot 10^{30}$ erg. Using these data, \cite{paudel2018k2} estimated the occurrence rate as a function of flare energy using a power law with the slope of $\alpha = 1.61$, which is consistent with other estimates \citep{Vida2017,ducrot2020trappist}. Assuming Kepler's observations are representative of the long-term activity of the star, we obtain $\bar{N}_f \approx 202$ flares per year in the aforementioned energy band. These observations serve as constraints for our long-term stochastic modeling. Given the number of events per unit time and their energy distribution we can simulate the flaring activity of the system over a chosen period of time by randomly sampling the power law distribution (Appendix \ref{app:gen_and_prop_CME}). Additionally, although flares with energies much higher than the maximum energy detected for TRAPPIST-1 are known to occur for other M dwarfs \citep{paudel2018k2}, we set an upper limit for flare energies to be $E_{max} = 10^{34}$ erg to avoid physically unrealistic or uncommon events. Therefore, all events with energies greater than $E_{max}$, which occur as we sample the energy distribution, are ignored. Our results justify this approach since the contribution of high-energy events to interior heating starts to decrease beyond a certain energy since such events become increasingly rare. 

\begin{figure}[ht!]
\centering
\includegraphics[width=0.6\textwidth]{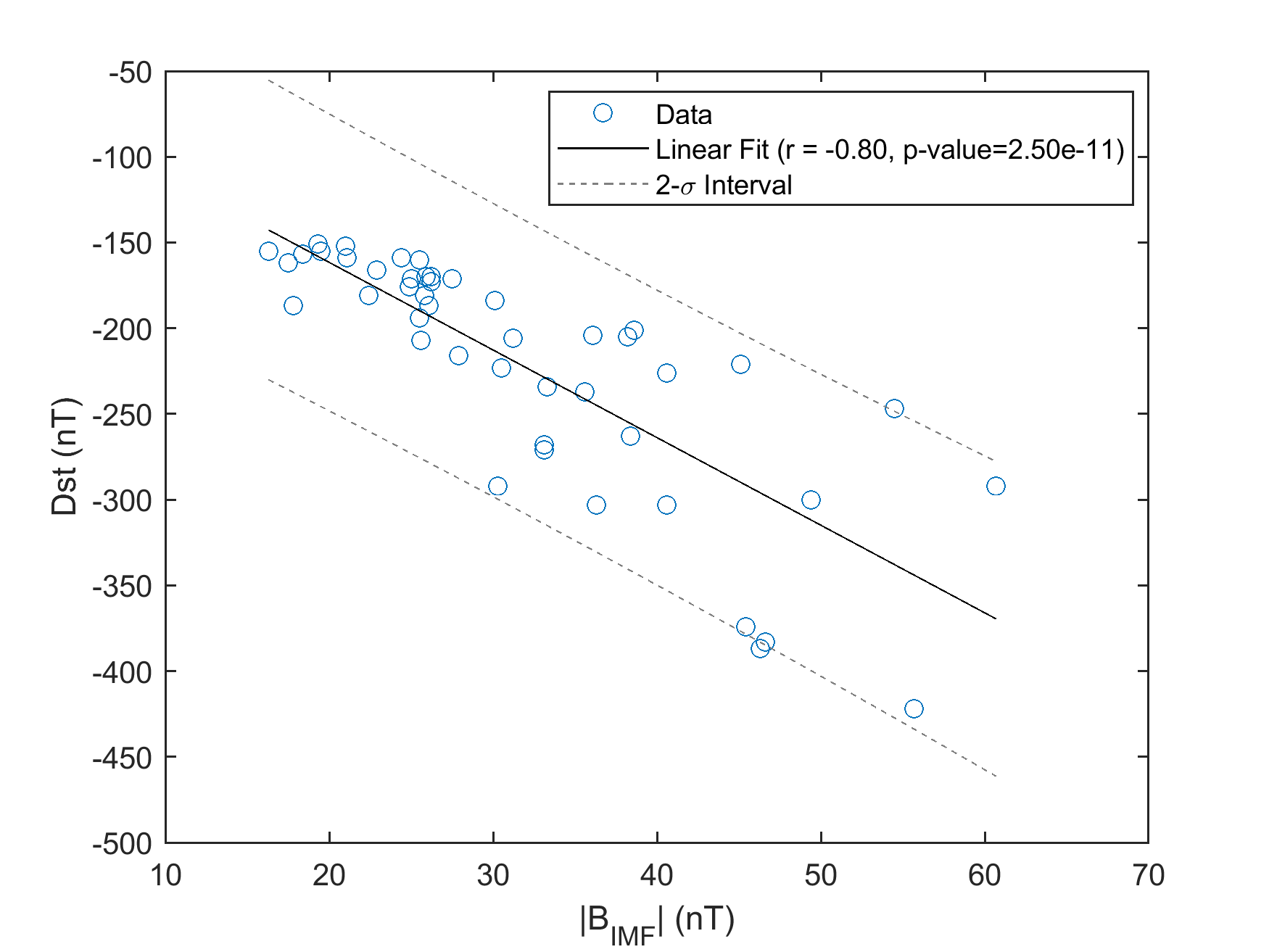}
\caption{Relation between the magnitude of the interplanetary magnetic field (IMF) and magnetic disturbance index (circles) as well as a linear fit (black solid line) and the corresponding uncertainty range (dashed lines). The regression line slope and intercept coefficients are $-5.1$ and $-59.4$, respectively. Data provided from the OMNI database. \label{fig:imf_vs_dst}}
\end{figure}

Randomly generated flares of a given energy are assumed to be associated with CME events. These events are propagated to the planet and the ICME magnetic field strength, $M_{ICME}$, is calculated following the flux-rope model of \cite{samara2021readily} (Appendix \ref{app:gen_and_prop_CME}). The peak external magnetic field strength at a planet is then obtained as
\begin{equation}
\label{eq:peak_perturbation}
M_{peak}(E_{flare},r_p) = 
\begin{cases}
  M_{ICME}(E_{flare},r_p) & \text{No intrinsic field} \\
  M_{Dst}(E_{flare},r_p) & \text{Earth-like intrinsic field}.
\end{cases}
\end{equation}
Thus, the external field strength is given by the ICME field if no intrinsic field exists. In presence of an Earth-like magnetic field, the peak external field strength in Eq. \ref{eq:peak_perturbation} is given by $M_{Dst}(E_{flare},r_p)$, which is inferred from the relation between the IMF magnitude ($|B_{IMF}|$) and the $Dst$ index (Figure \ref{fig:imf_vs_dst}). This relation was constructed using the OMNI database \citep{king2005solar} for geomagnetic storms with $Dst < -150$ nT that occurred between years 1996 and 2021. Assuming that upon collision with an ICME, interplanetary magnetic field is mostly due to the ICME magnetic flux, we take $|B_{IMF}| \equiv M_{ICME}(E_{flare},r_p)$. The linear fit model shows that upon interaction of the Earth's magnetosphere with an ICME, the magnitude of effective external magnetic field perturbations exerted on the Earth is larger than the IMF by a factor of $\approx 3-9$ \citep{akasofu1981energy}.

Since not every ICME will strike a planet, we need to account for the collision frequency of an ICME with a planet. The frequency of an impact is calculated following the model of \citep{khodachenko2007coronal} (see Appendix \ref{app:gen_and_prop_CME} for more details). The model predicts that, on average, only 8\% of flare-associated ICMEs will interact with a planet. We refer to ICMEs that actually collide with a planet as `effective'. Thus, among all events sampled from the flare frequency distribution (see Appendix \ref{app:gen_and_prop_CME}), we randomly select a fraction of effective ICMEs according to Eq. (\ref{eq:prob_impact}). Since all TRAPPIST planets have different orbital periods and ICMEs propagate at a finite speed, the selection of effective flares is performed for each planet independently.

We determine the inductive response of a planet due to temporal changes in the external magnetic field associated with the passage of ICMEs.  A system of governing partial differential equations, namely Ampere's and Faraday's laws (see Appendix \ref{app:induction} for a detailed derivation) are solved to determine the distribution of electric currents induced in the interior by an external forcing.  Although some interplanetary magnetic field persists as a result of stellar wind flow in the interplanetary space \citep{baumjohann1996basic}, induction is mostly driven by the rate of change of magnetic field \citep{parkinson1983}. Therefore, we only model times when a flare-associated ICME collides with a planet since temporal changes in external field are largest during these events.  Large background magnetic field variations would further enhance the total heat dissipated in the interior \citep{kislyakova2017magma}.

In addition to external magnetic field variations, the magnitude of the induced electric currents, and thus the Joule heating, have a nonlinear dependence on the interior electrical conductivity. Although the physical model of EM induction within a rocky planet (Appendix \ref{app:induction}) allows one to choose an arbitrary $\sigma(\vec{r})$, there are no constraints on the lateral conductivity heterogeneity in rocky exoplanets. Therefore, we set $\sigma(\vec{r}) \equiv \sigma(r)$, that is to first order the conductivity varies only in the radial direction due to pressure effects and phase transitions in minerals. We adopt the radial profile that is characteristic of the Earth-like pyrolitic mantle \citep{grayver2017joint} and is scaled according to the planet size (Figure \ref{fig:profiles}a). Following \cite{agol2021refining}, the average core-mantle boundary (CMB) is assumed to be $0.48R$, where $R$ is the radius of the planet.  A constant conductivity of $10^6$ S/m is appropriate for the iron core beneath the silicate mantle. Heat generated in the core is excluded from consideration because it represents a negligible fraction of the total heat since electric currents attenuate exponentially \citep{parkinson1983} within the electrically conducting model of the mantle. In addition to an Earth-like radial conductivity profile, we also tested a range of radially homogeneous conductivity models (Figure \ref{fig:profiles}b,c).

The amount of Joule heating within the planet was then calculated as described in Appendix \ref{app:heating}. Specifically, we calculate the time-averaged volumetric heat energy:
\begin{equation}
\label{eq:heat_avg}
    Q_{avg} = \frac{1}{t_2 - t_1}\int_{t_1}^{t_2}\int_{V} \left[\mathbf{E} \cdot \mathbf{J}\right] \textnormal{d}V\textnormal{d}t,
\end{equation}
where the integrated electric field $\mathbf{E}$ and electric current density $\mathbf{J}$ are functions of space and time. To get an intuition into the main terms and parameters that affect the dissipated energy in planets, the reader is referred to Appendix \ref{app:heating}, where the closed form analytical solution for a conducting sphere is derived. The averaging time interval, $[t_1, t_2]$, should be sufficiently long such that it contains statistically representative samples of effective flares. In this study, the time interval of one year was used. Within each time interval, the average heat energy was calculated using a randomly sampled set of effective flares that conform to the flare frequency distribution (FFD) of the TRAPPIST-1 star \citep{paudel2018k2}. Furthermore, to account for uncertainty and variability in input parameters, we defined some of them as random variables. Table \ref{tab:params} lists all randomly sampled parameters along with their prior probability distributions. 

\section{Results} 
\label{sec:results}

\begin{figure}
\gridline{\fig{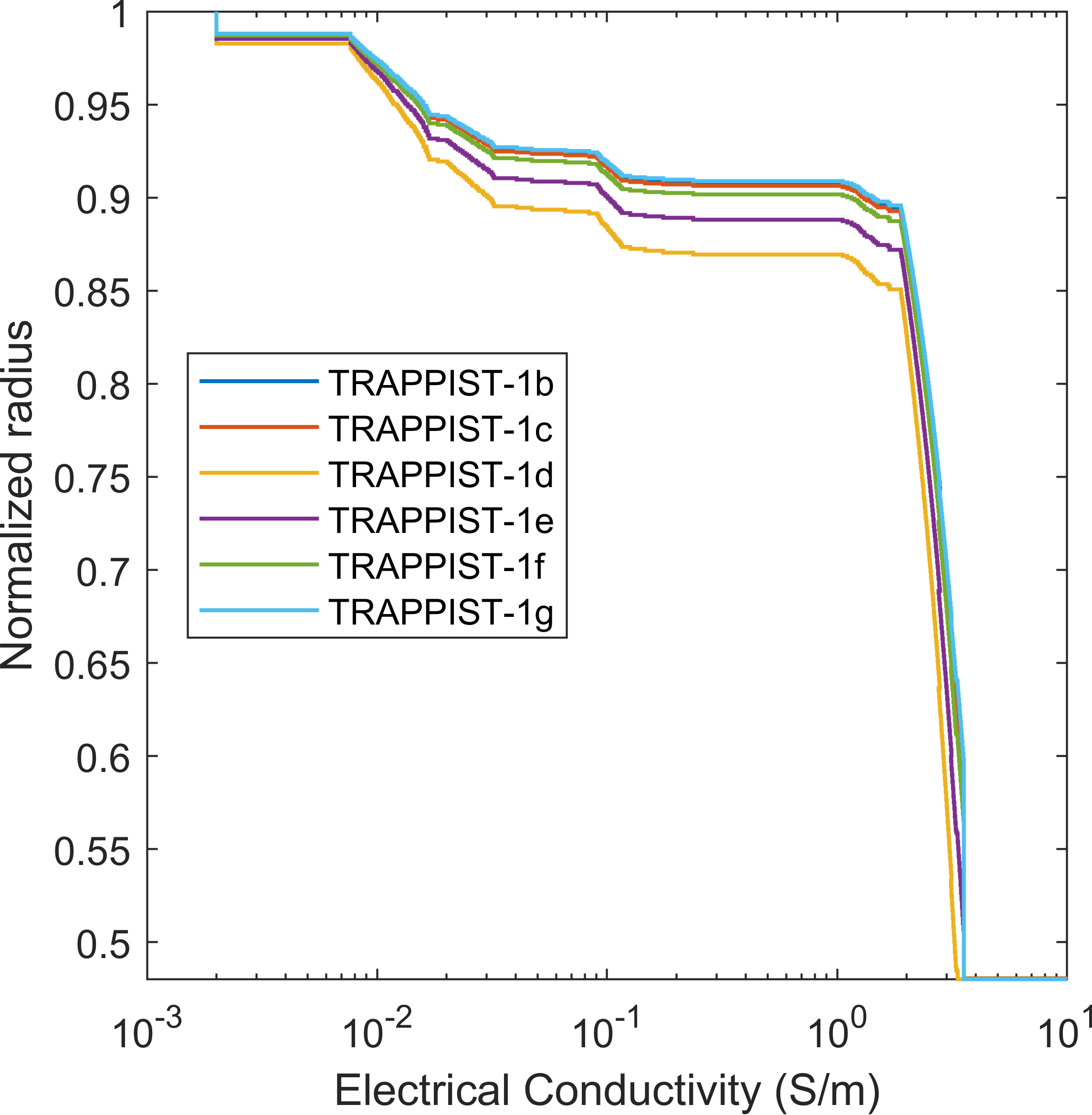}{0.3\textwidth}{(a)}
          \fig{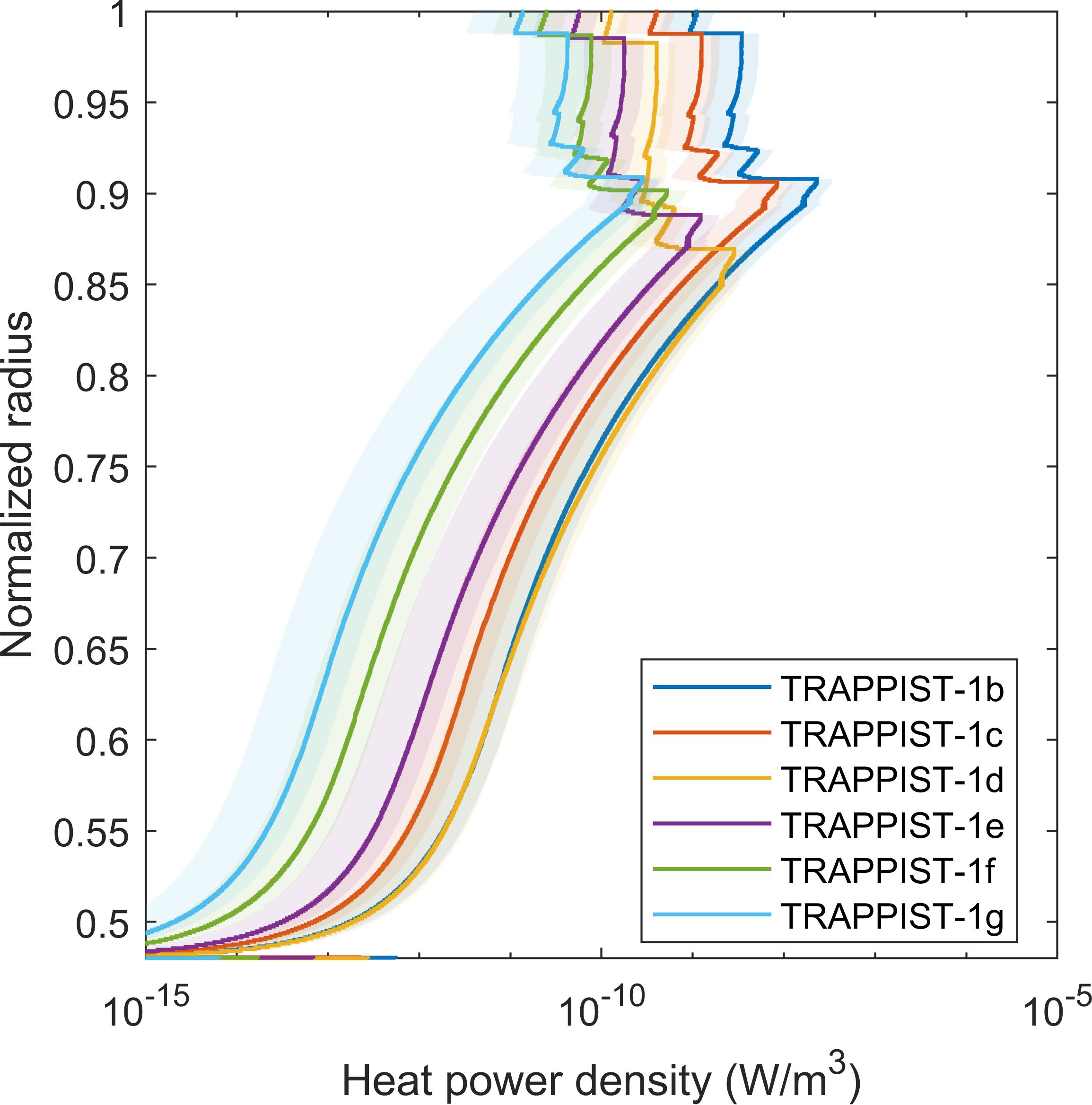}{0.3\textwidth}{(d)}
          \fig{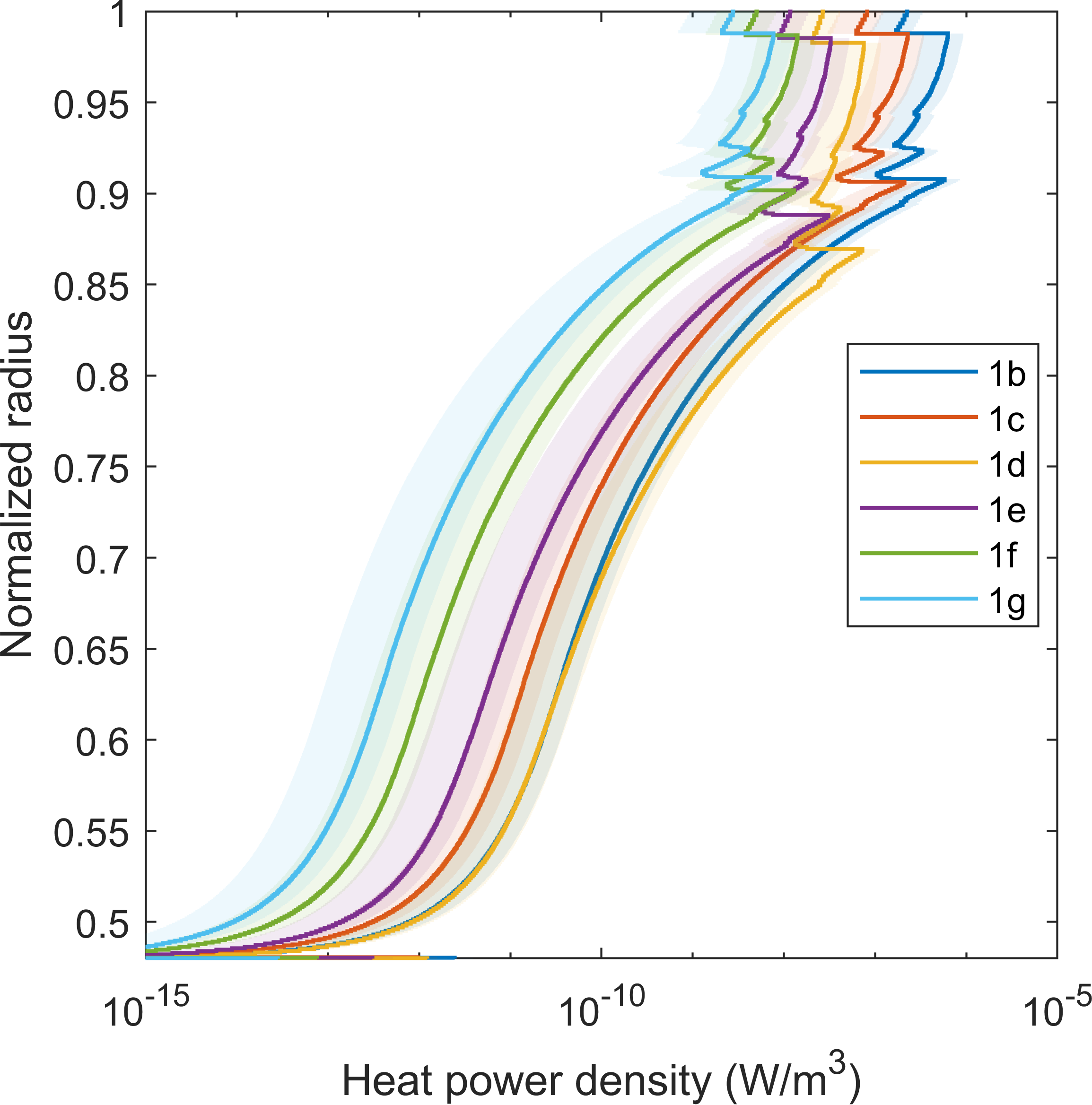}{0.3\textwidth}{(g)}
          }
\gridline{\fig{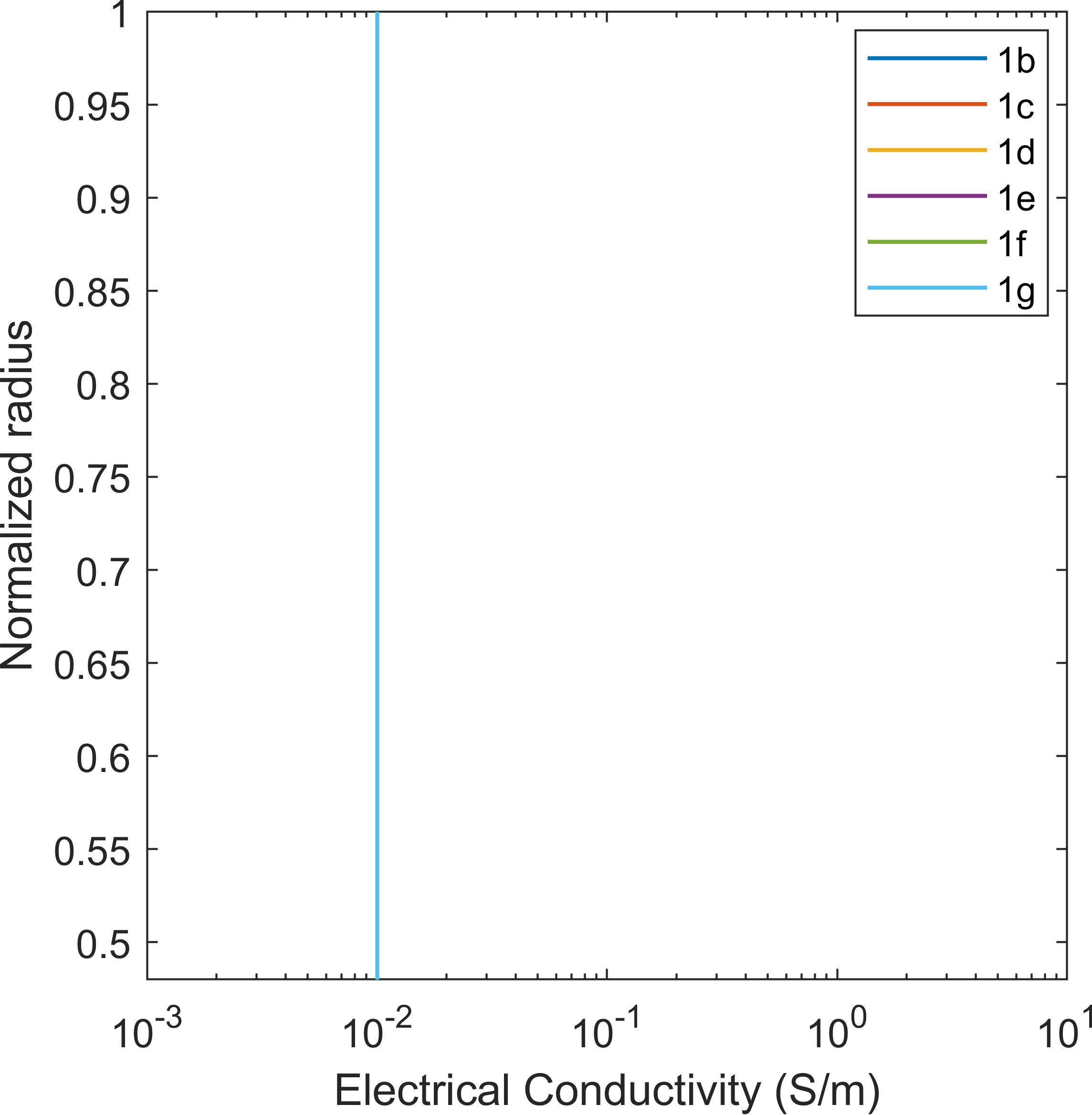}{0.3\textwidth}{(b)}
          \fig{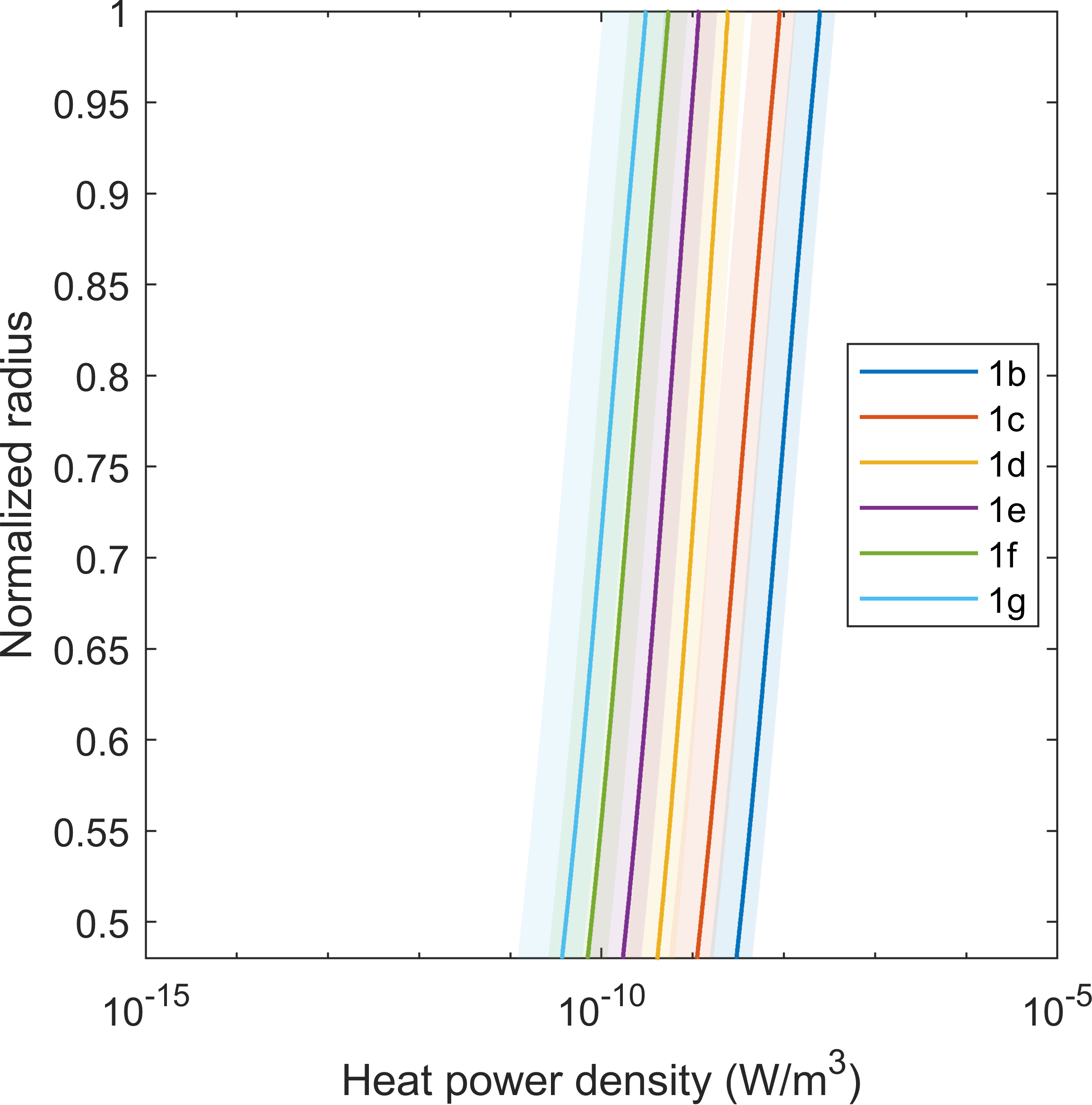}{0.3\textwidth}{(e)}
          \fig{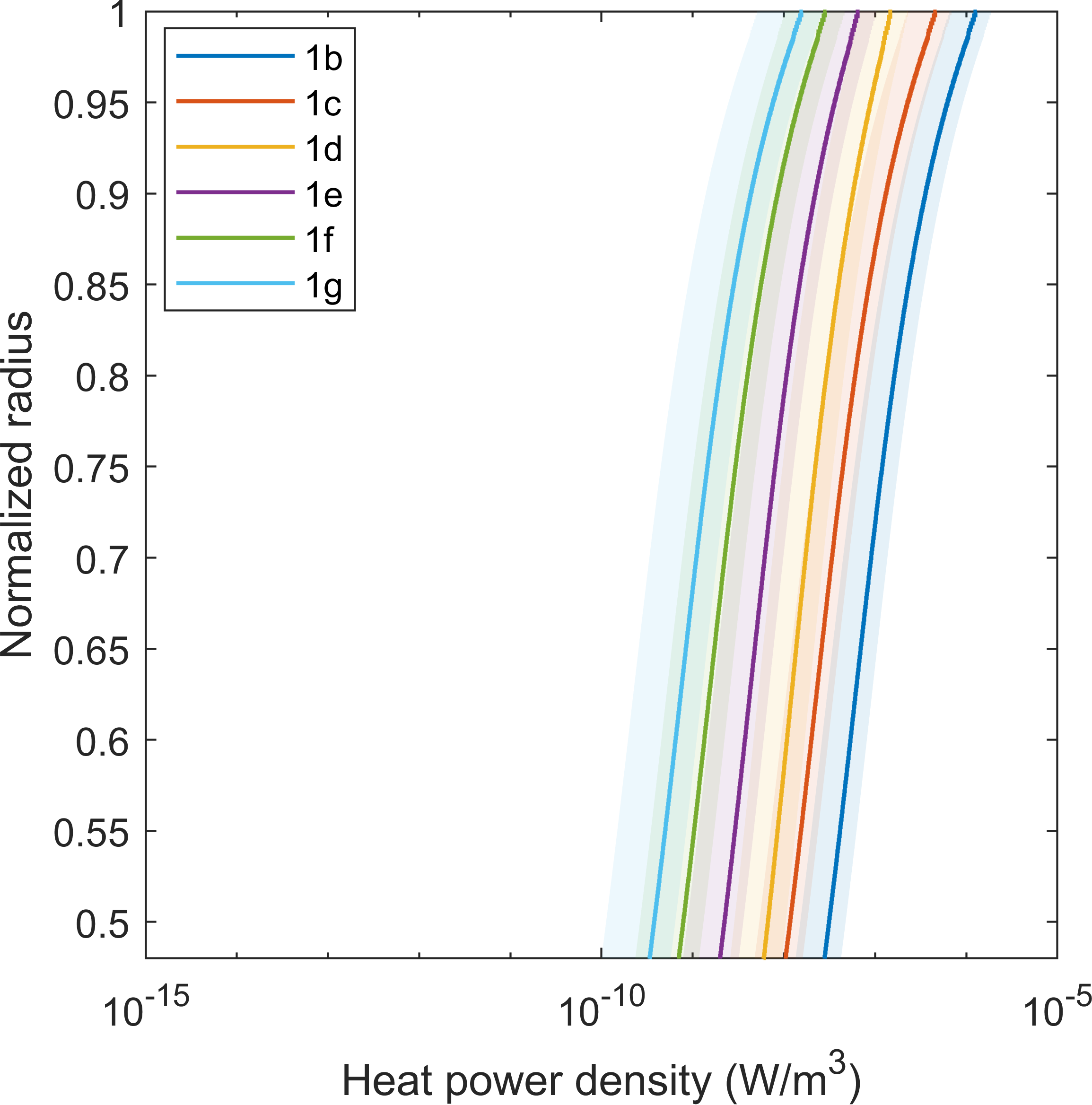}{0.3\textwidth}{(h)}
          }
\gridline{\fig{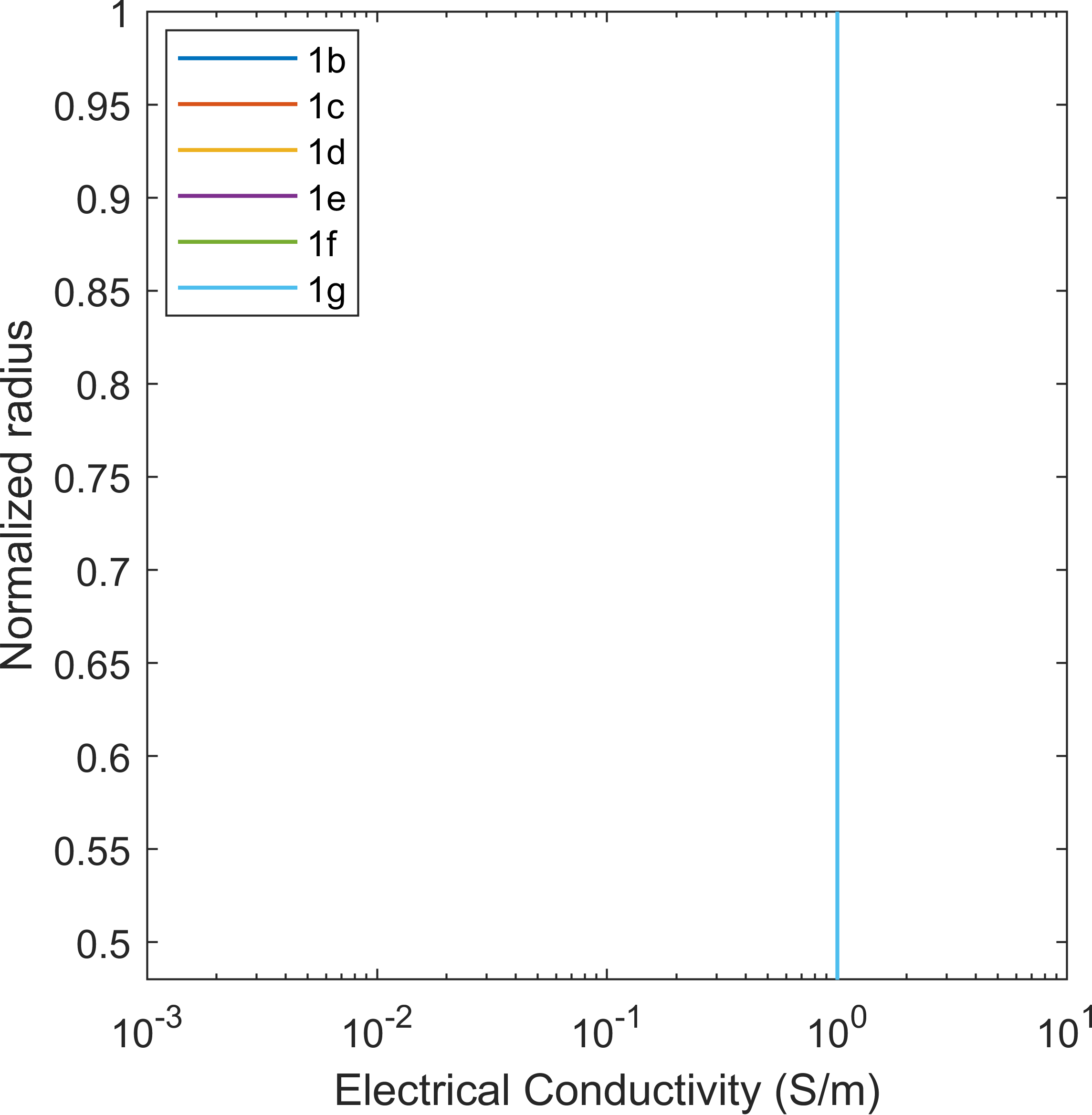}{0.3\textwidth}{(c)}
          \fig{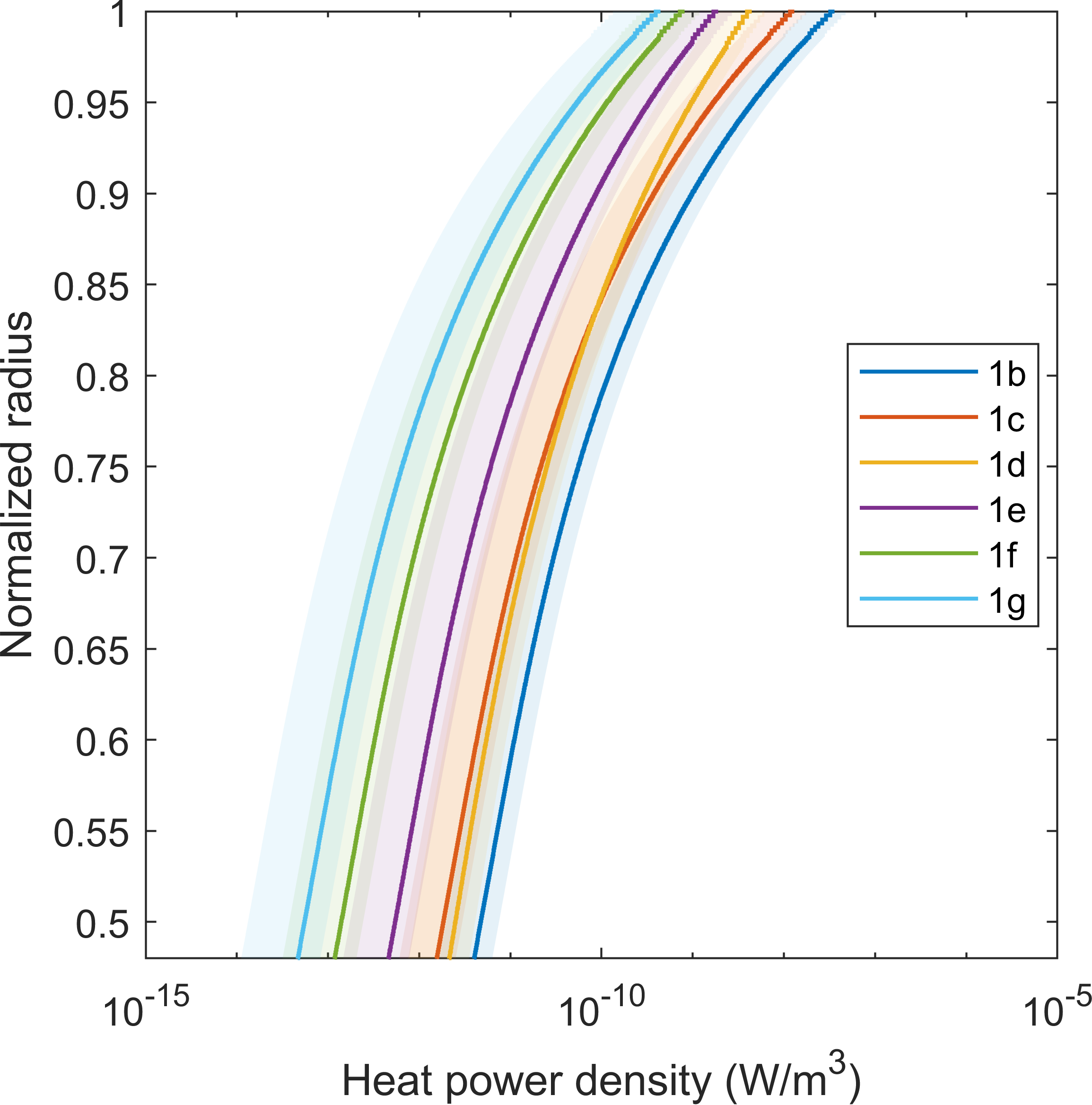}{0.3\textwidth}{(f)}
          \fig{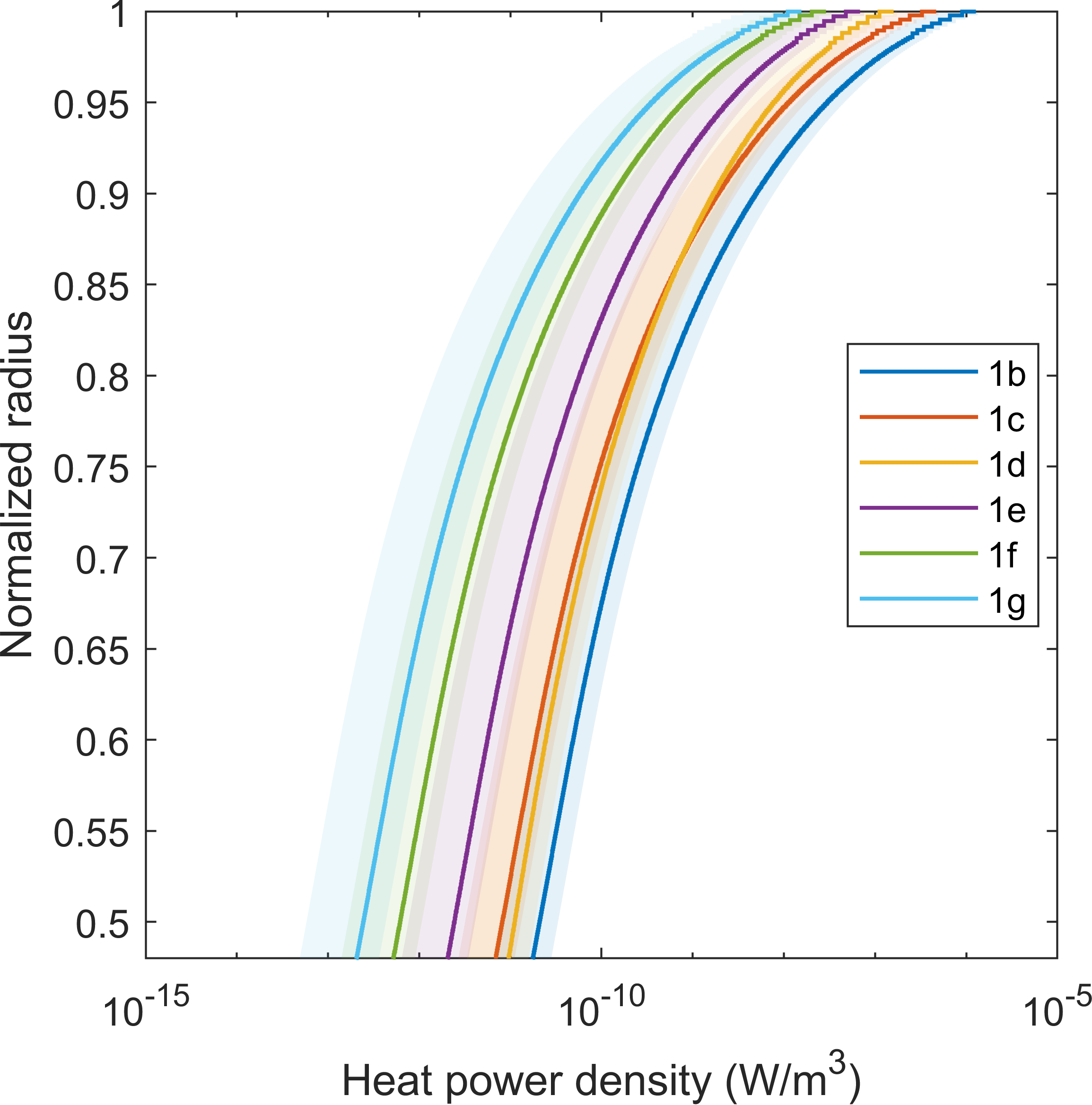}{0.3\textwidth}{(i)}
          }
\caption{Interior heating of the TRAPPIST-1 planets 1b--1g due to flare-associated CMEs.  (a-c) Electrical conductivity profiles (note in case of the homogeneous conductivity models, profiles for different planets overlap). (d-f) Mean profiles of the time-averaged Joule heating assuming no intrinsic planetary magnetic field. (g-i) Same as (d-f), but with an Earth-like magnetic field. Shaded areas bound one standard deviation.  For the constant conductivity models, 0.01 S/m is characteristic of peridotite at upper mantle conditions \citep{yoshino2013electrical} and 1 S/m roughly corresponds to a 5\% partial silicate melt at 1600 K.
\label{fig:profiles}}
\end{figure}

A stochastic simulation was run for 50,000 years, which is sufficient to exhaustively sample the observed flare energy distribution and all varied parameters (Table \ref{tab:params}). We used one year as a time interval to calculate the time-averaged heat energy as per Eq. (\ref{eq:heat_avg}). The average annual number of effective flares was around 16, resulting in more than 700,000 events for each planet during the entire simulation and enabling us to calculate statistics. Once a `burn-in' phase was passed, which typically lasts a few thousand years, the amount of dissipated heat depends solely on flare frequency distribution of the star, model of interplanetary CME propagation, frequency of a collision, and the interior conductivity profile.  We recall that some variables (Table \ref{tab:params}) that determine the aforementioned parameters were also randomly sampled for every year of the simulation. 

Figure~\ref{fig:profiles} shows estimates of the heat power density due to ohmic dissipation throughout the planetary mantles of the TRAPPIST-1 planets. It shows radial profiles of the time-averaged heat power density for cases with and without a planetary magnetic field and several subsurface conductivity profiles. The shaded areas bound one standard deviation of the ensemble of heating profiles that are calculated over the entire simulation.  The Joule heating decreases with the distance from the star as the IMF decays upon propagation (Appendix \ref{app:gen_and_prop_CME}), and it also attenuates with depth in the planet as a result of the skin-depth effect \citep{parkinson1983}. The attenuation with depth depends on the planet radius and is larger for a more conductive interior. This effectively concentrates the majority of the dissipated heat in the uppermost part of the planetary mantle ($\approx 600$ km). We also observe that for a 1-D pyrolitic conductivity model, the heat deposition has local maxima at depths of large conductivity jumps, which may facilitate melting in these regions. In general, the presence of (partially) molten layers, which are much more conductive than solid phases \citep{yoshino2013electrical}, results in higher local heating because the current density is proportional to the electrical conductivity (Eq.~\ref{eq:heat_avg} and Appendix \ref{app:induction}). However, the attenuation of the EM field is higher within a conductive medium. Therefore, the total amount of heat dissipated within partially molten layers will depend on their thickness and lateral extent.  Figure~\ref{fig:profiles}c shows a homogeneous conductivity model of 1 S/m corresponding to a partially molten silicate material. According to a basalt melt model \citep{laumonier2017experimental}, this is approximately equivalent to a 5\% melt fraction at $1600$ K.

\begin{figure}
\centering{NO PLANETARY MAGNETIC FIELD}
\gridline{\fig{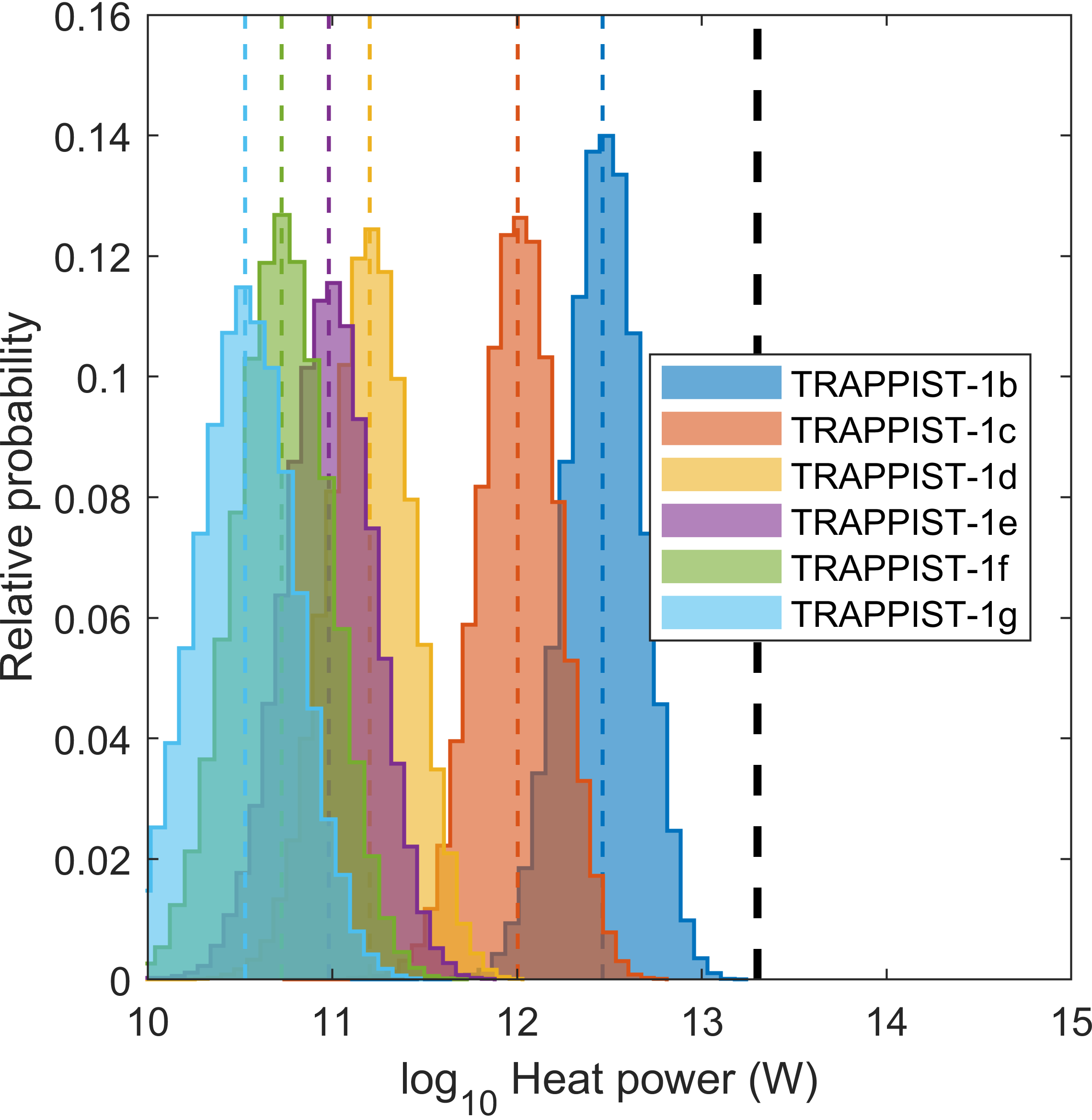}{0.25\textwidth}{(a)}
          \fig{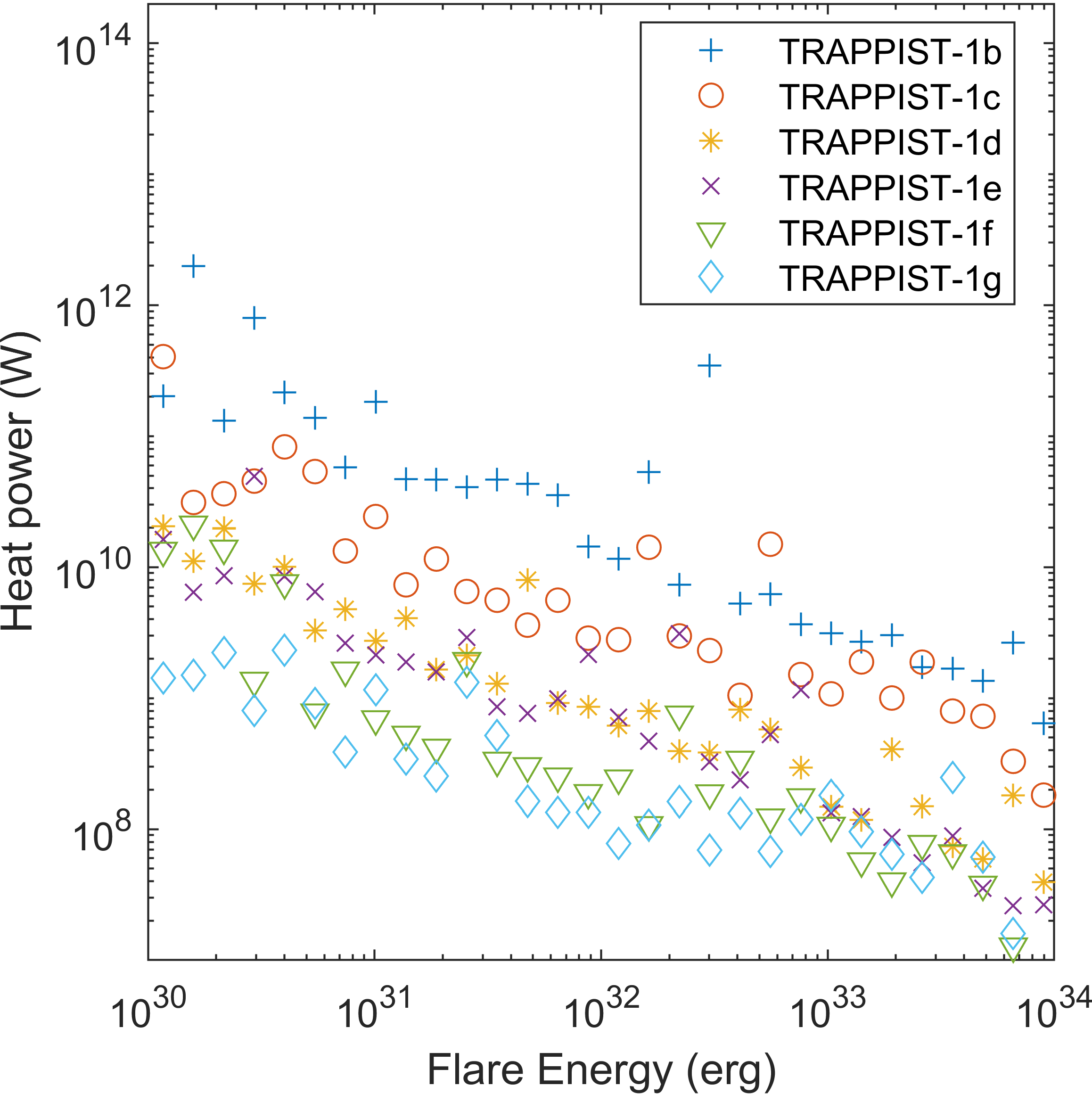}{0.25\textwidth}{(b)}
          \fig{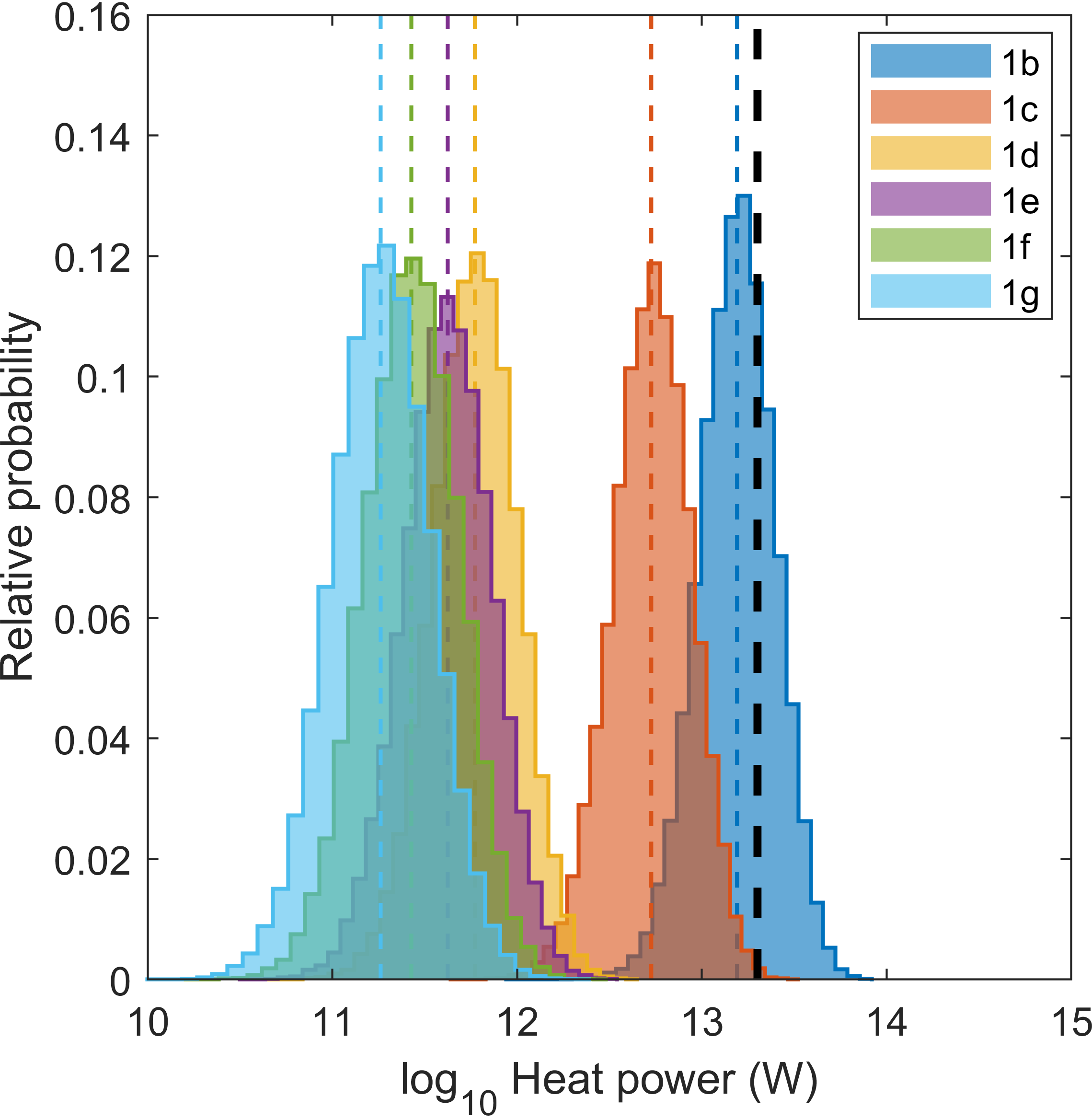}{0.25\textwidth}{(c)}
          \fig{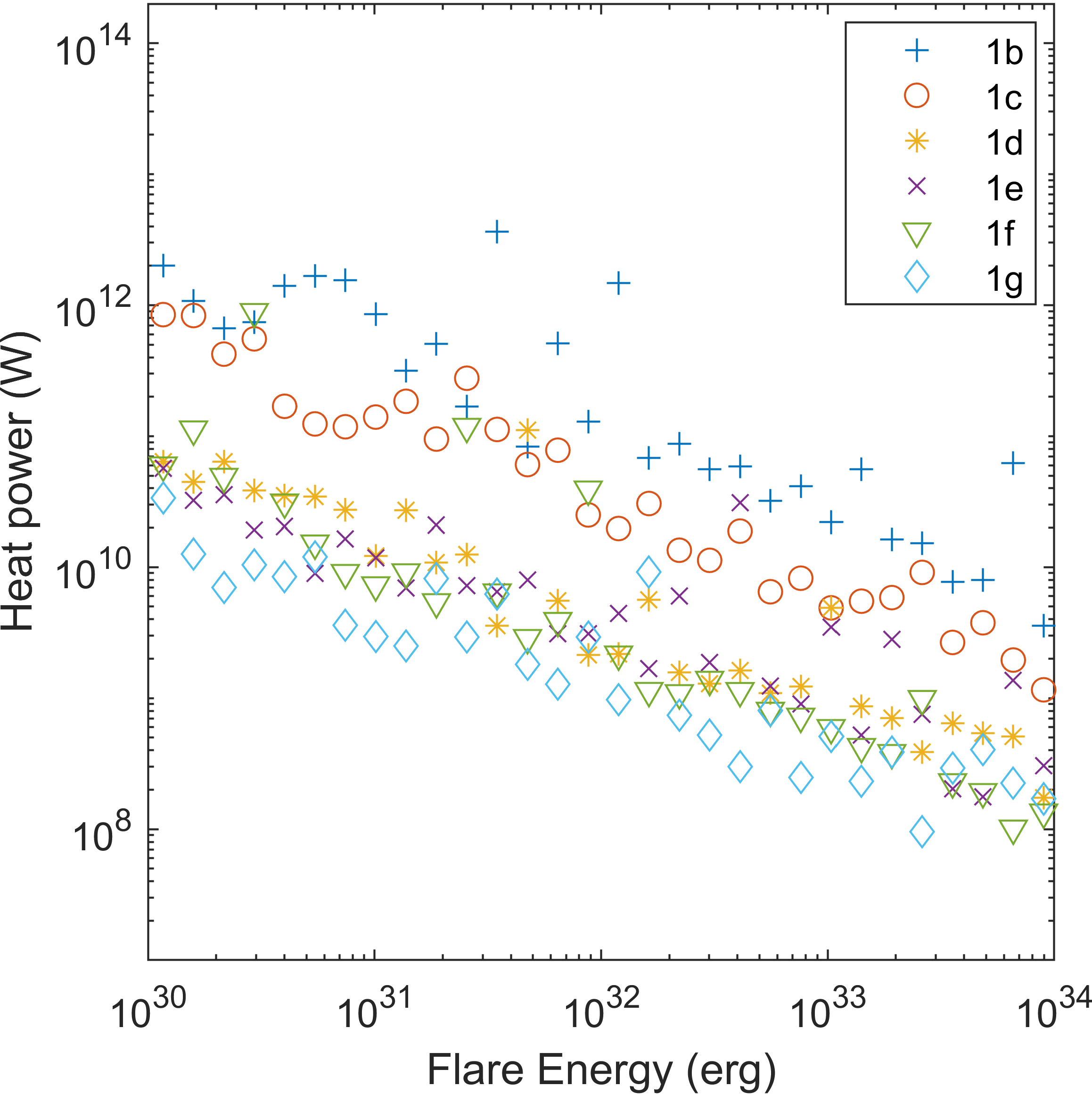}{0.25\textwidth}{(d)}
          }
\centering{WITH INTRINSIC MAGNETIC FIELD}
\gridline{\fig{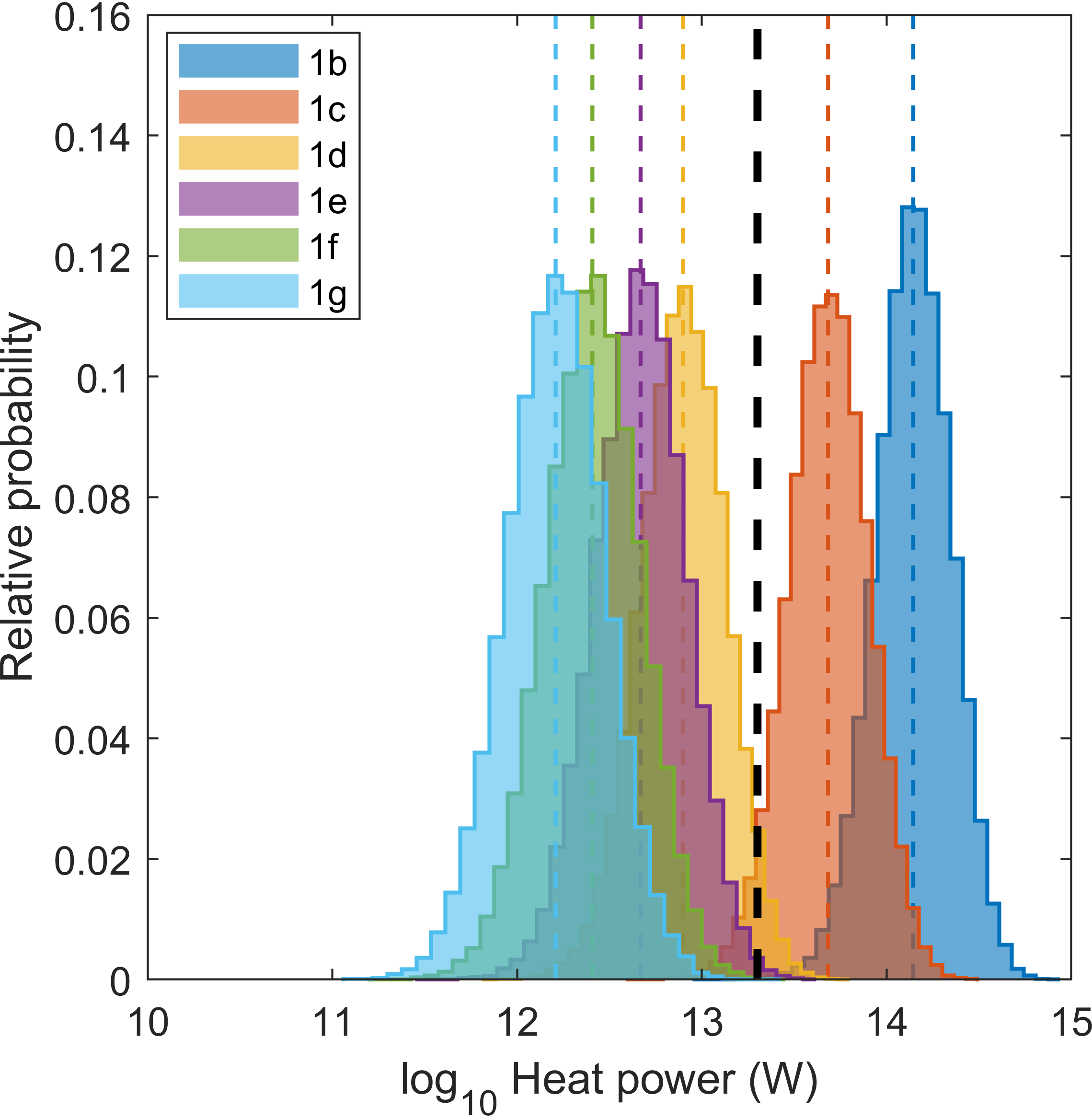}{0.25\textwidth}{(e)}
          \fig{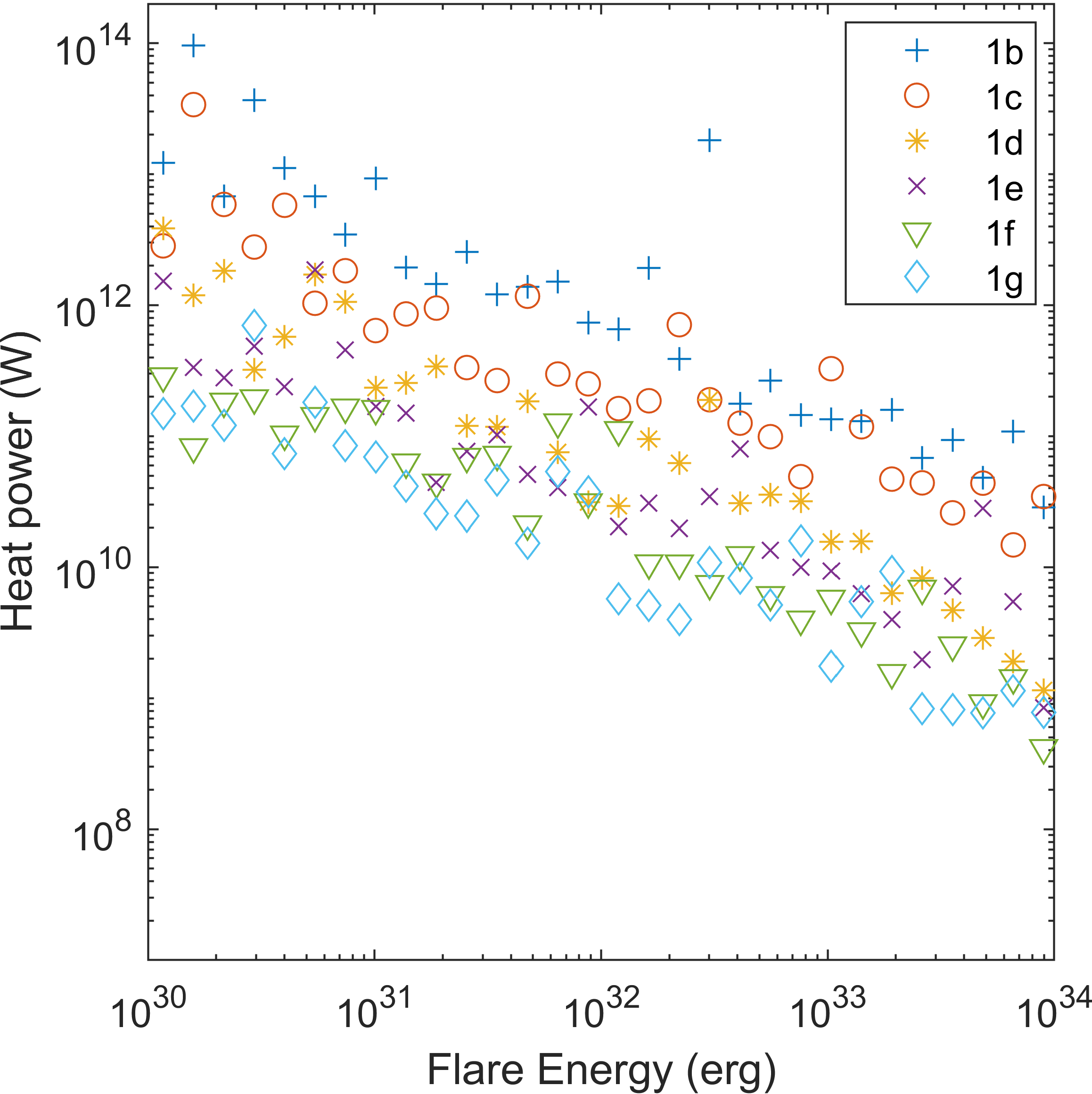}{0.25\textwidth}{(f)}
          \fig{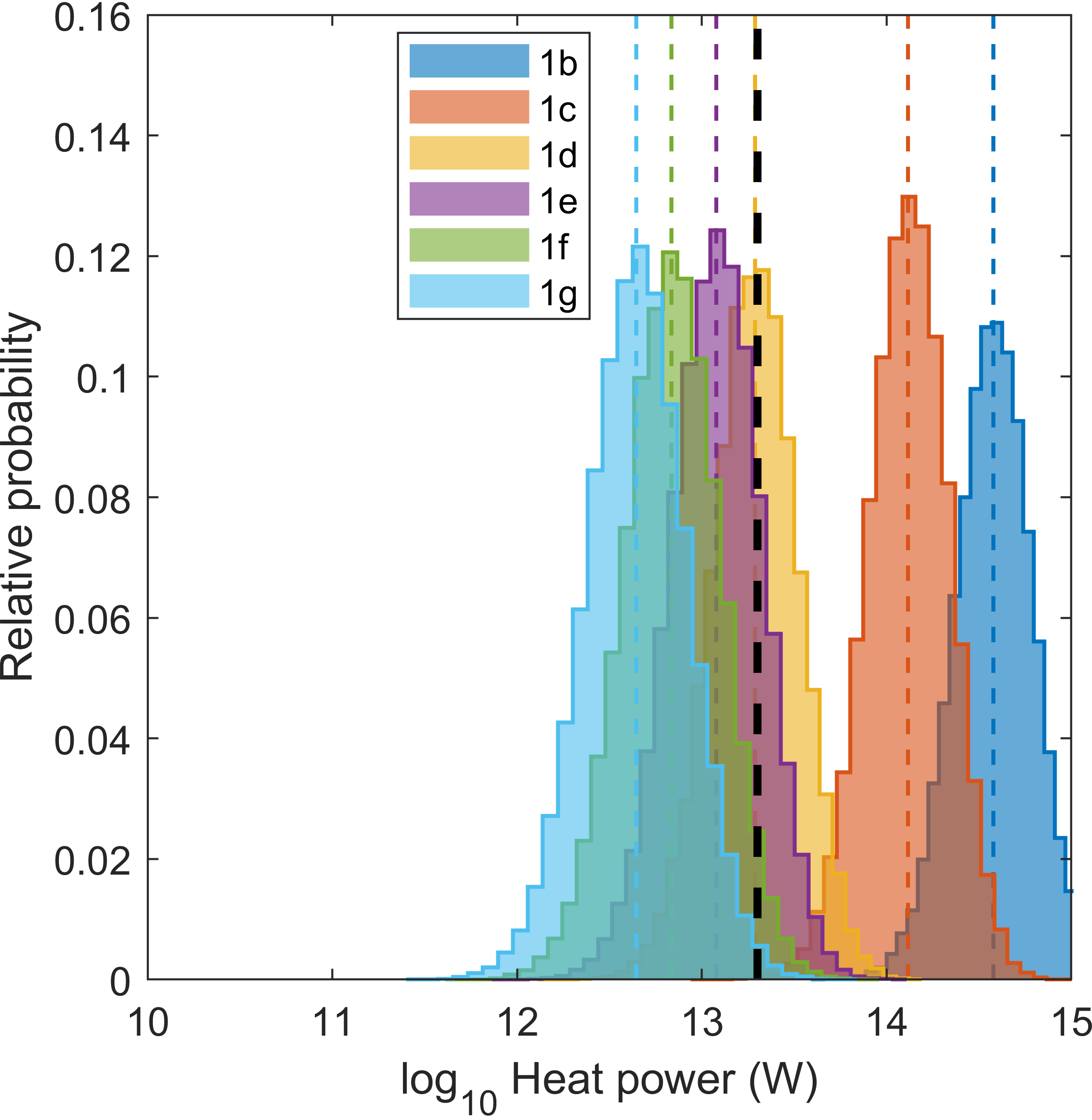}{0.25\textwidth}{(g)}
          \fig{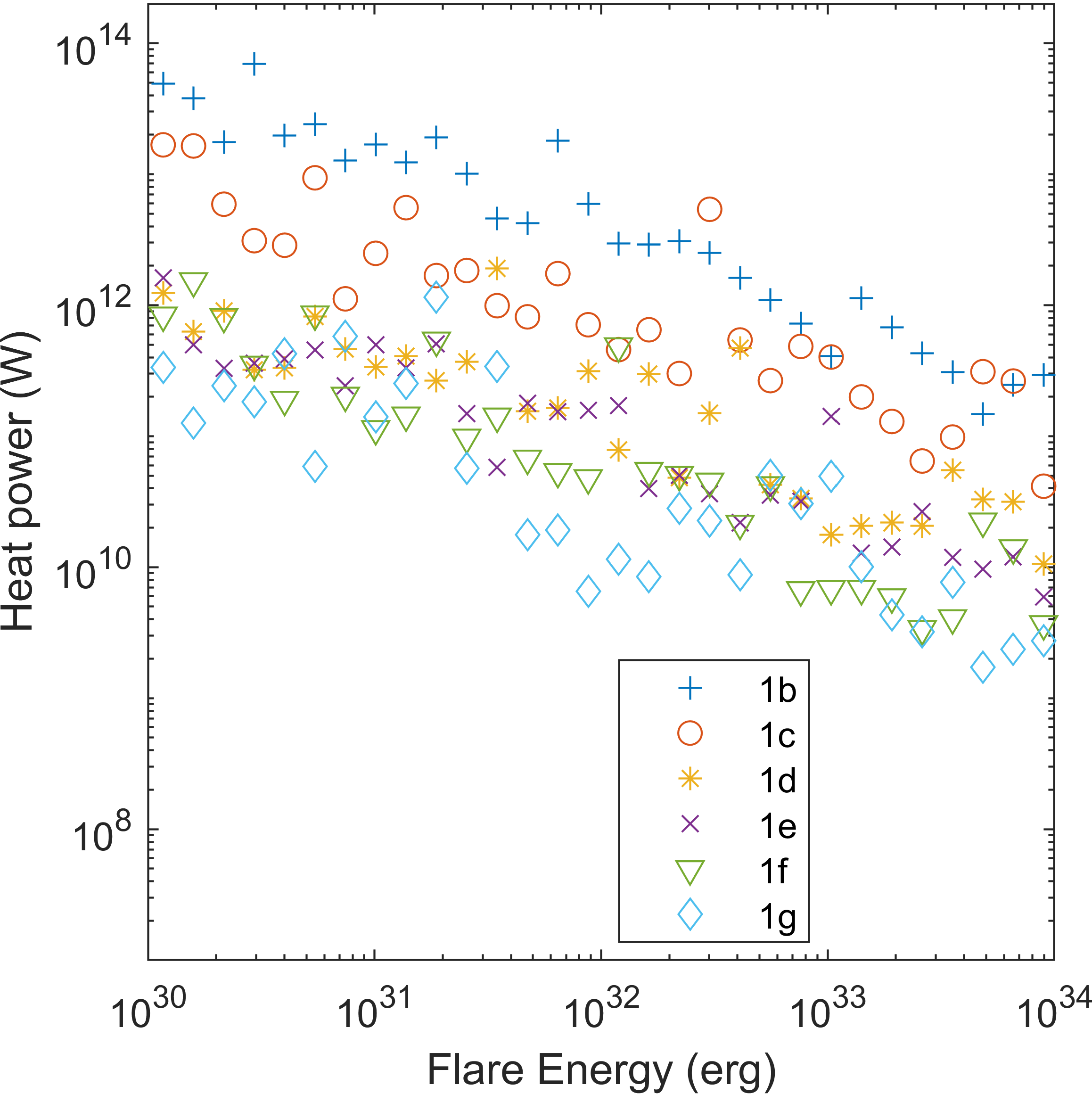}{0.25\textwidth}{(h)}
          }
\caption{(a): Normalized probability distributions of the average volumetric heating rate generated in each TRAPPIST-1 planet by flare-associated ICMEs, for the 1-D radial conductivity model and no planetary magnetic field. Colored dotted lines correspond to geometric mean values and the thick dashed line delineates 20 TW which is Earth's present-day radiogenic heat production. (b) Average volumetric heat produced by all flares within a specified energy interval for a planet without an intrinsic magnetic field. Markers denote the center of the interval. (c-d) Same as (a-b), but for a constant 0.01 S/m conductivity model. (e-h) Same as (a-d), but with an Earth-like magnetic field.
\label{fig:hist}}
\end{figure}

Figure \ref{fig:hist} shows histograms of the total volumetric heating rate for each planet along with the distribution of dissipated heat as a function of the flare energy. The variability in the dissipated heat stems from the annual variations in randomly perturbed parameters (Table \ref{tab:params}), resulting in well-shaped lognormal distributions. Another clear observation is the dependency of the heating rate on flare energy. Despite some scatter resulting from the stochastic nature of the models, it is evident that less energetic events are able to produce more heat over geological timescales because they outnumber more rare high energy events. This emphasizes the importance of considering both the energy spectrum and flare frequency. Both Figures \ref{fig:profiles} and \ref{fig:hist} reveal that more heat is generated internally for planets that possess an Earth-like magnetic field.  This is because interactions of the incoming ICME magnetic field with a magnetosphere lead to larger variations in the external (inducing) magnetic field \citep{akasofu1981energy, gopalswamy2008solar} impinging on the planet's surface. 

\begin{table}[]
\centering
\caption{Comparison of heating rates (in TW) for different heating mechanisms. Electromagnetic heating was calculated for the layered (Figure \ref{fig:profiles}a) and a homogeneous (100 $\Omega$m) models.}
\label{tab:comparison}
\begin{tabular}{|l|llll|ll|l|}
\hline
\multicolumn{1}{|c|}{\multirow{3}{*}{Planet}} & \multicolumn{4}{c|}{Ohmic (ICMEs)$^1$} & \multicolumn{2}{c|}{Ohmic (stellar field)$^{2}$} & \multicolumn{1}{c|}{\multirow{3}{*}{Tidal$^{3}$}} \\ \cline{2-7}
\multicolumn{1}{|c|}{} & \multicolumn{2}{c|}{Layered} & \multicolumn{2}{c|}{Homogeneous} & \multicolumn{1}{c|}{\multirow{2}{*}{Layered}} & \multicolumn{1}{c|}{\multirow{2}{*}{Homogeneous}} & \multicolumn{1}{c|}{} \\ \cline{2-5}
\multicolumn{1}{|c|}{} & \multicolumn{1}{l|}{Nonmagnetized} & \multicolumn{1}{l|}{Magnetized} & \multicolumn{1}{l|}{Nonmagnetized} & Magnetized & \multicolumn{1}{c|}{} & \multicolumn{1}{c|}{} & \multicolumn{1}{c|}{} \\ \hline
b & \multicolumn{1}{l|}{2.9} & \multicolumn{1}{l|}{139.2} & \multicolumn{1}{l|}{15.5} & 379.3 & \multicolumn{1}{l|}{42.9} & 304.5 & N/A \\ \hline
c & \multicolumn{1}{l|}{1} & \multicolumn{1}{l|}{48.2} & \multicolumn{1}{l|}{5.3} & 130.5 & \multicolumn{1}{l|}{4.5} & 19.1 & N/A \\ \hline
d & \multicolumn{1}{l|}{0.16} & \multicolumn{1}{l|}{7.9} & \multicolumn{1}{l|}{0.59} & 19.4 & \multicolumn{1}{l|}{0.34} & 0.3 & N/A \\ \hline
e & \multicolumn{1}{l|}{0.1} & \multicolumn{1}{l|}{4.6} & \multicolumn{1}{l|}{0.42} & 11.9 & \multicolumn{1}{l|}{0.31} & 1.1 & 12.2 \\ \hline
f & \multicolumn{1}{l|}{0.05} & \multicolumn{1}{l|}{2.5} & \multicolumn{1}{l|}{0.26} & 6.8 & \multicolumn{1}{l|}{0.18} & 1. & 17.0 \\ \hline
h & \multicolumn{1}{l|}{0.034} & \multicolumn{1}{l|}{1.6} & \multicolumn{1}{l|}{0.18} & 4.4 & \multicolumn{1}{l|}{0.11} & 0.73 & 0.72 \\ \hline
\end{tabular}
\\
{$^1$ The reported values are geometric means as shown in Figure \ref{fig:hist}}\\
{$^2$ Values were calculated following the approach of \cite{kislyakova2017magma}. Note that we used the updated TRAPPIST-1 rotation period of 3.3 days, whereas in \cite{kislyakova2017magma} an older estimate of 1.4 days was used. Other parameters are identical.}\\
{$^3$ Estimates from the work of \cite[][Table 3]{bolmont2020solid}. Layer-averaged model was used and we calculated mean values for different eccentricities assumed in the original study. We note that the layer-averaged models in \cite{bolmont2020solid} are not strictly speaking compatible with our conductivity models.}
\end{table}

\section{Discussion and Conclusions}
\label{sec:discussion}

Based on the available data on the flaring activity of TRAPPIST-1, the heat generated inside the TRAPPIST-1 planets could be comparable to the energy released by Earth's radionuclides at present-day ($\approx 20$ TW), particularly if the planet possesses an intrinsic magnetic field and/or is close to the star (Figure \ref{fig:hist}).  For planets without a magnetic field, 20 TW should be considered an upper bound, except for perhaps planet 1b where 20 TW falls near the center of the distribution for the constant 0.01 S/m conductivity profile. Table \ref{tab:comparison} compares dissipated heat produced by different mechanisms. We see that heat produced by the motion through a periodically varying stellar magnetic field \citep{kislyakova2017magma} is larger than the mean heat due to ICMEs if planets are nonmagnetized and smaller in case of magnetized planets. The tidal heating estimates from \citep{bolmont2020solid} are larger for all reported outer planets, although in case of the magnetized planets, the estimates are comparable. We note that electric currents, and hence Joule heating, will generally attenuate with depth, as a result of the skin-depth effect \citep{parkinson1983}. Consequently, electromagnetic heating is the most significant in the uppermost part of a planet. The exact distribution of heating depends on external forcing and electrical conductivity, but the general trends we find are robust given the positive correlation between temperature (which increases with depth) and conductivity for major mineral phases \citep{yoshino2013electrical}.  Figure \ref{fig:profiles} reveals that the maximum heating rate is observed in the uppermost 300--400 km. Because of this, local heating rates can be much more significant despite relatively small total volumetric values.

The presence or not of an intrinsic planetary magnetic field strongly dictates how significantly the interior is heated (Figure \ref{fig:hist}).  Assuming an Earth-like dipolar field, interaction of stellar plasma and an intrinsic magnetosphere increases temporal variations of external magnetic perturbations exerted on a planet; this results in higher current density within the planet.  Hence while intrinsic planetary magnetic fields mitigate the impact of plasma particles on planetary atmospheres, at the same time they amplify external magnetic field variations which leads to larger amounts of heat produced in the interior.  Induced magnetospheres can also form around planets without dynamo fields, as with Mars and Venus \citep{ramstad2021intrinsic}, but their amplifying effect, if any, will be much smaller compared to intrinsic magnetospheres, and the peak magnetic field variations exerted on a planet will mostly be driven by magnetic energy carried within an ICME \citep{samara2021readily}. Presently there is no information on the existence or strength of an intrinsic magnetic field for any TRAPPIST-1 planets. In the solar system, Earth possesses the strongest dynamo field among rocky objects; the fields of Mercury and Ganymede are much weaker although they are also significantly smaller and their rotation rate is slower. Hence our assumption of an Earth-like magnetic field could be viewed as an upper estimate for the TRAPPIST-1 planets. In addition, we used a $Dst$-based proxy to calculate the amplification of the external field variations exerted on a planet with an Earth-like field. On one hand, $Dst$ index accounts only for an amplification in the first zonal harmonic of the field, hence missing variations from nonaxisymmetric components of magnetospheric currents and polar current systems, which are very significant during magnetic storms \citep{finlay2017challenges}. Therefore, the effective amplification is higher for the Earth and the $Dst$-based proxy represents a conservative estimate. On the other hand, translating this proxy to all TRAPPIST-1 planets does not account for differences in local plasma and stellar wind environments. In particular for close-in planets, this might result in smaller magnetospheric cross-sectional areas for the stellar wind to interact with \citep[e.g.][]{fischer2019}.  Depending on the magnetic field strengths of the Trappist-1 planets and their ambient stellar wind magnetic field and ram pressures, the amplification could be reduced for the closer-in planets. Thus, a $Dst$-based proxy should be considered an approach well established for Earth-like environments, and further studies exploring the wider parameter space underlying the interactions between the Trappist-1 planets and their surrounding stellar wind is required.

We choose to apply our model to the TRAPPIST-1 system, but the list of stars with observed flaring activity and planets in the HZ is growing \citep{paudel2018k2, Seli2021, ilin2021giant}.  Hence our model can be readily used to study other planetary systems as well.  Flare-induced heating will be most prominent for smaller flaring stars with HZ very close to the star ($\approx 0.1$ AU), which is the case for stars of spectral type M. For larger stars with the HZ significantly farther than $\approx 0.1$ AU, such as solar-type stars, the effect of stellar flares on interior heating will be greatly reduced due to decay of the ICME energy with distance (Appendix \ref{app:gen_and_prop_CME}).  Close-in planets, such as TRAPPIST-1b and 1c, possibly lie within the stars' Alfven radius, such that they can in principle couple electromagnetically to the star and no bow shock evolves upstream of the planets, establishing a unique form of energy exchange (see \citep{fischer2019} and references therein).  If established, such close coupling between a star and planet could change the external magnetic field variations experienced by the planet during a CME event, and hence the heating rate could be different from our estimates. Implementation of close-in stellar-wind interactions is an avenue for future research.

Recent simulation studies suggest that CMEs occurring on strongly magnetized stars have a difficulty leaving the corona and remain either partially or fully confined. A study that specifically targeted active M dwarf stars \citep{alvarado2019coronal} suggests that the TRAPPIST-1 star would be in a weakly/moderately confining regime if one adopts the most probable field strength value of 600 G \citep{Reiners2010}.  This would result in fewer CMEs that actually reach the planets, especially CMEs with lower energy.  However, this conclusion depends on many other parameters that enter MHD simulations, including the geometry of the stellar field, which is presently unknown for TRAPPIST-1. Additionally, given that our model is stochastic and the number of ICME events is varied (Table~\ref{tab:params}), we partially account for this uncertainty in the amount of ICME events.  Furthermore, some studies suggest that flares on active late-type stars predominantly occur at high latitudes \citep[$55^\circ-81^\circ$,][]{ilin2021giant}, thus planets with orbits in the equatorial plane may not experience the full impact of associated ICMEs even if they collide. However, our probabilistic impact model (Eq. \ref{eq:prob_impact}) assumes that flare CMEs occur in the latitude range of $\pm 80^{\circ}$, thus covering the relevant range reported by \cite{ilin2021giant} and consistent with uniformly distributed surface magnetic activity \citep[e.g.][]{Feinstein2020}. In addition, in the presence of a stellar dipolar field, high-latitude CMEs have a tendency to be deflected toward the equatorial plane upon their propagation \citep{kay2019frequency}, rendering even polar CMEs potential candidates for collision with planets that reside in the stellar ecliptic plane.


A future extension to our model would be to accommodate time-dependent orbital and stellar evolution scenarios. For instance, changes in stellar flaring activity will adjust the amount of dissipated heat within planets. Equivalently, changes in the orbital configuration that affect orbit axes or inclination will lead to different radial heating profiles. Another temporal effect is the change of interior electrical conductivity as a result of temperature changes associated with heating. However, as an ICME collides with a planet, the majority of the heat will be produced in the following few days or weeks (Figure \ref{fig:field_vs_heat}b), depending on the impact duration and the electrical conductivity of the subsurface. In contrast, tangible variations in conductivity, stellar flaring activity or orbital characteristics, will occur on much longer timescales. Therefore, the essentially instantaneous dissipation of heat due to a collision with an ICME will be decoupled in time from other long-term temporal changes in stellar activity or orbital configuration.

Stellar magnetic dynamos are ultimately powered by stellar rotation, and M dwarfs spin down inefficiently compared to their FGK siblings, over timescales of billions of years \citep[e.g.][]{Newton2018,Curtis2019}. Flaring and other magnetic activity could be expected to persist over geological timescales (Gyr), as the decline of magnetic activity with age becomes quite weak for late-Ms \citep{Reiners2010,Seli2021}.  Hence immaterial of other heat sources, flare-induced heating can provide sufficient energy over Gyrs for geological processes to sculpt the surface and atmosphere of the planet, such as by tectonic processes, volcanic resurfacing, and outgassing.  This is because maintaining a hot interior (compared to a cooler surface temperature) drives interior convection and hence cooling, and ultimately it is this energy that powers geological processes. Short-lived radionuclides such as Al26 are only effective at shaping early planetary building blocks and the early magma ocean phase of rocky planet evolution is dominated by secular cooling.  Tidal heating is most relevant for close-in planets with non-negligible eccentricity, which effectively requires planet--planet interactions to prevent orbit circularization otherwise tidal heating is also short-lived.  Therefore, the energy to power geological processes is derived from long-lived interior heat sources (like long-lived radionuclides) as well as residual heat from formation.  Here we demonstrate that for active stars with close-in planets, such as the TRAPPIST-1 system, interior heating due to flare activity could also be a significant contributor to the interior heat budget of planets over long timescales.  Nevertheless, from the perspective of planetary habitability, the potentially destructive role of flares/ICMEs on the atmosphere and surface could mitigate the beneficial role they play in powering a geologically active planet.

We apply  boundary layer theory to estimate the increase in the average interior temperature of the TRAPPIST-1 planets due to average flare heating for a 1-D layered conductivity model compared to radiogenic heating alone ($\Delta T$, Appendix~\ref{app:boundarylayer}, Table~\ref{tab:boundarylayer})o. If the planets do not possess an intrinsic magnetic field, the increase in the mantle temperature is up to 100 K, which will not fundamentally change the global mantle dynamics although it could accentuate local dynamics through increased melt production.  Most strikingly, however, is the interior temperature for the innermost planets if they possess an Earth-like magnetic field. For planet b the mantle temperature can be elevated by more than a 1000 K, which is more than sufficient to extensively melt certain regions of the mantle to produce large melt reservoirs or magma oceans. The preferential deposition of heat in the uppermost mantle could facilitate the formation or sustenance of molten regions that sequester incompatible elements (volatiles) enabling their delivery to the surface and subsequent outgassing to the atmosphere. This outgassing could rejuvenate the atmosphere if some or all of the previous atmosphere was stripped away by its direct interaction with ICMEs. In the presence of melt there are several feedback mechanisms that become important.  First, the higher conductivity of partial melt compared to solid (Figure~\ref{fig:profiles}) would increase the attenuation rate, limiting the ability of ohmic heating to sustain melt at depths once it had formed.  Second, melt will migrate according to the local pressure gradient (Darcy's law), thus melt would be redistributed from regions where it formed. Coupling with tidal heating would introduce further complexity since tidal heating depends on the multilayered structure of the planet.  Therefore, including flare-induced heating (along with other heat sources) in coupled interior--atmosphere simulations will be necessary to assess in detail how flares impact interior--surface--atmosphere interactions. 

Our major findings are as follows.  First, using observational data of the flare frequency distribution of TRAPPIST-1 we find that particularly the innermost planets (1b and 1c) can experience an interior heating rate comparable to Earth's current radiogenic heat production due to ICMEs alone.  Second, the heating rate is enhanced in the presence of an intrinsic planetary magnetic field, particularly for the innermost planets that would experience an interior temperature increase of up to a thousand K. We note that it remains unknown if the TRAPPIST planets have magnetic fields. Third, more frequent lower energy ICMEs can result in more interior heating than infrequent high energy ICMEs. This contrasts with atmospheric stripping models which are particularly dependent on high energy events.  Fourth, for stellar dynamos that are sustained by stellar rotation, flare activity can persist for Gyr, and thereby provide an interior heat source of both magnitude and duration comparable to long-lived radionuclides.  Therefore, ICME-induced heating alone can sustain a hot interior over long timescales to enable geological processes to occur (volcanism, outgassing), even in the absence of long-lived radiogenics. Our study is therefore directly relevant for ongoing searches for atmospheres on rocky exoplanets. A dedicated JWST campaign, for example, is scheduled to observe TRAPPIST-1c in November 2022 with the aim to discover and constrain an atmosphere on this planet \citep{2021jwst.prop.2304K}.

\begin{acknowledgments}
All data used in this study are publicly available. The OMNI data were obtained from the GSFC/SPDF OMNIWeb interface at \url{https://omniweb.gsfc.nasa.gov}. Data and scripts to reproduce results from figures from Zenodo at \url{https://doi.org/10.5281/zenodo.7262947}. AG was supported by the Heisenberg Grant from the German Research Foundation, Deutsche Forschungsgemeinschaft (Project No. 465486300). DJB carried out this work within the framework of the NCCR PlanetS supported by the Swiss National Science Foundation under grants 51NF40\_182901 and 51NF40\_205606. JS is supported by the European Research Council (ERC) under the European Union’s Horizon 2020 research and innovation programme (grant agreement No. 884711). We are thankful to the reviewer, whose insightful comments helped us improve the original work.
\end{acknowledgments}






\appendix

\section{Electromagnetic induction in rocky planets}
\label{app:induction}

\noindent Distribution of electric and magnetic fields within a rocky planet is governed by Maxwell's equations:
\begin{eqnarray}
\mu^{-1}\nabla \times \mathbf{B} &=& \mathbf{J}, \label{eq:maxwell_time1} \\
\nabla \times \mathbf{E} &=& -\frac{\partial\mathbf{B}}{\partial t}, \label{eq:maxwell_time2}
\end{eqnarray}
where $\mu \equiv \mu_0 = 4\pi \times 10^{-7}$ [H/m] is the magnetic permeability of free space; $\mathbf{B}$ [T], $\mathbf{E}$ [V/m] are magnetic and electric fields, respectively.  Vector $\vec{r}=(r, \vartheta, \varphi)$ describes a position in a planet-fixed reference frame with $r$, $\vartheta$ and $\varphi$ being radial distance, co-latitude, and longitude, respectively. The current density, $\mathbf{J}$ [A/m$^2$], is
\begin{equation}
    \mathbf{J} = \mathbf{J}^c + \mathbf{J}^{\text{ext}}, \label{eq:current_density}
\end{equation}
where $\mathbf{J}^c = \sigma\mathbf{E}$ denotes conduction current and $\mathbf{J}^{\text{ext}}$ is a primary (external) source current density; $\sigma(\vec{r})$ [S/m] is electrical conductivity. Displacement currents were omitted from the equation \ref{eq:current_density} due to their negligible effect in the low frequency regime we consider \citep{parkinson1983}. Additionally, we assume a linear medium where $\mathbf{B} = \mu\mathbf{H}$, where $\mathbf{H}$ is the magnetic field intensity [A/m]. For rocky planets considered this study, we assumed $\mu \equiv \mu_0$ throughout the volume, although this assumption may not be justified for iron-rich bodies at low temperatures \citep{bromley2019ohmic}. Adopting the Fourier convention
\begin{equation}
f(t)=\frac{1}{2\pi}\int\limits_{-\infty}^{\infty}\tilde{f}(\omega)e^{\mathrm{i}\omega t}\mathrm{d}\omega
\label{eq:fourier}
\end{equation}
allows us to rewrite equations (\ref{eq:maxwell_time1}-\ref{eq:maxwell_time2}) in the frequency domain as
\begin{eqnarray}
\mu^{-1}\nabla \times \mathbf{\tilde B} &=& \mathbf{\tilde J}, \label{eq:maxwell_freq1} \\
\nabla \times \mathbf{\tilde E} &=& -\textrm{i} \omega \mathbf{\tilde B}, \label{eq:maxwell_freq2} 
\end{eqnarray}
where $\omega$ denotes the angular frequency. All vector quantities defined above are functions of space ($\vec{r}$) and/or time (respectively, frequency). These dependencies are implied, but omitted for brevity. 

We assume that the source currents $\mathbf{\tilde J}^{\text{ext}}(\vec{r},\omega)$ flow above the solid surface of the planet. Specifically, if $R$ is the radius of a planet, there is a value $b > R$ such that $\mathbf{\tilde J}^{\text{ext}}(\vec{r},\omega) \neq 0$ only for $r \ge b$, while the region $r \in (R, b)$ is an insulator. This allows us to represent any external current density distribution using an equivalent current system given by
\begin{eqnarray}
\mathbf{\tilde J}^{\text{ext}}(\vec{r}, \omega) &=& \sum_{n = 0}^{N}\sum_{m = -n}^{n} \mathbf{J}_n^m(\vec{r}) \tilde{\varepsilon}_n^m(\omega),
\label{eq:current}
\end{eqnarray}
with
\begin{equation}
\label{eq:jnm}
\mathbf{J}_n^m(\vec{r}) = \frac{\delta(r-b)}{\mu_0} \frac{2n+1}{n+1}\left(\frac{b}{R}\right)^{n-1} \hat{e}_r \times \nabla_{\perp} S_n^m(\theta, \phi),
\end{equation}
where $b = R + h$ and $h$ is the altitude of the current sheet. Further, 
\begin{equation}
S_n^m(\theta, \phi) = P_n^{|m|}(\cos{\theta})\exp{(\textnormal{i}m\phi)}
\end{equation}
is a spherical harmonic (SH) function of degree $n$ and order $m$ with $P_n^{|m|}$ being Schmidt semi-normalized associated Legendre polynomials \citep{parkinson1983} with corresponding complex-valued SH coefficients $\tilde{\varepsilon}_n^m(\omega)$ [T]. Finally we have
\begin{equation}
\label{eq:nabla_perp}
\nabla_{\perp} = \frac{\partial}{\partial\theta}\hat{e}_{\theta} + \frac{1}{\sin{\theta}}\frac{\partial }{\partial\phi}\hat{e}_{\phi}
\end{equation}
with $\hat{e}_r$, $\hat{e}_{\theta}$ and $\hat{e}_{\phi}$ being unit vectors of the body-fixed spherical coordinate system.  The current density (Eq.~\ref{eq:current}) will result in magnetic field variations, $\mathbf{\tilde B}^{\text{ext}}(\vec{r}, \omega)$, of external origin to the planet. These variations can be reproduced exactly anywhere in region $R \leq r < b$ using a scalar potential
\begin{eqnarray}
\label{eq:magpotential_ext}
  \mathbf{\tilde B}^{\text{ext}}(\vec{r},\omega) = -\nabla \tilde{V}^{\text{ext}}(\vec{r}, \omega),
\end{eqnarray}
where
\begin{eqnarray}
\label{eq:magpotential_ext_fd}
  \tilde{V}^{\text{ext}}(\vec{r}, \omega) = R \sum_{n=1}^{N}\sum_{m = -n}^{n} \tilde{\varepsilon}_n^m(\omega) \left( \frac{r}{R} \right)^n S_n^m(\theta, \phi).
\end{eqnarray}

Employing linearity of Maxwell's equations with respect to the $\mathbf{J}^{\text{ext}}$ term, the total EM field within a planet can be expanded as
\begin{eqnarray}
\mathbf{\tilde B}(\vec{r}, \omega; \sigma) &=& \sum_{n=1}^{N}\sum_{m = -n}^{n}\mathbf{\tilde B}_n^m(\vec{r}, \omega; \sigma)  \tilde{\varepsilon}_n^m(\omega) \label{eq:b_field_sum}, \\
\mathbf{\tilde E}(\vec{r}, \omega; \sigma) &=& \sum_{n=1}^{N}\sum_{m = -n}^{n}\mathbf{\tilde E}_n^m(\vec{r}, \omega; \sigma)  \tilde{\varepsilon}_n^m(\omega) \label{eq:e_field_sum},
\end{eqnarray}
where $\mathbf{\tilde B}_n^m$ and $\mathbf{\tilde E}_n^m$ are unit current fields given as solutions of the following equations:
\begin{eqnarray}
\mu^{-1}\nabla \times \mathbf{\tilde B}_n^m &=& \sigma \mathbf{\tilde E}_n^m + \mathbf{\tilde J}_n^m \label{eq:maxwell_mode_1},\\
\nabla \times \mathbf{\tilde E}_n^m &=& - \textrm{i} \omega \mathbf{\tilde B}_n^m \label{eq:maxwell_mode_2}.
\end{eqnarray}

Although this derivation has been performed in the frequency domain, their time domain counterparts are readily obtained by applying the inverse of Eq. (\ref{eq:fourier}). For instance, Eq. (\ref{eq:magpotential_ext_fd}) in the time domain is
\begin{eqnarray}
  V^{\text{ext}}(\vec{r}, t) &=& \textnormal{Re}\left\{R \sum_{n=1}^{N}\sum_{m = -n}^{n} \varepsilon_n^m(t) \left( \frac{r}{R} \right)^n S_n^m(\theta, \phi) \right\} \nonumber \\
  &=& R\sum_{n = 1}^{N}\sum_{m = 0}^n \left[ q_n^m(t)\cos(m\phi) + s_n^m(t)\sin(m\phi) \right]\left( \frac{r}{R} \right)^n P_n^m(\cos{\theta}),
\end{eqnarray}
where complex and real SH coefficients are related by
\begin{equation}
\label{eq:complex_real_sh}
\varepsilon_n^m = 
\begin{cases}
\frac{q_n^m - \textnormal{i}s_n^m}{2}, & m > 0 \\
\frac{q_n^{|m|} + \textnormal{i}s_n^{|m|}}{2}, & m < 0 \\
q_n^m, & m = 0
\end{cases}.
\end{equation}

Equations (\ref{eq:b_field_sum}-\ref{eq:e_field_sum}) become convolution integrals upon their transformation to time domain. For demonstration, we consider the simplest case of a uniform external magnetic field aligned with the vertical axis in the body-fixed frame, which can be described by a single SH function with $n = 1$ and $m = 0$. Thus, in the time domain, eqs. (\ref{eq:b_field_sum}-\ref{eq:e_field_sum}) reduce to
\begin{eqnarray}
\mathbf{B}(\vec{r}, t; \sigma) &=& \int_{-\infty}^t \mathbf{B}_1^0(\vec{r}, t - \eta; \sigma)  q_1^0(\eta)\textnormal{d}\eta, \label{eq:b_field_td} \\
\mathbf{E}(\vec{r}, t; \sigma) &=& \int_{-\infty}^t \mathbf{E}_1^0(\vec{r}, t - \eta; \sigma)  q_1^0(\eta)\textnormal{d}\eta, \label{eq:e_field_td}
\end{eqnarray}
Efficient calculation and properties of these integrals for a general case are discussed in \cite{grayver2021}.  Further, when $\sigma(\vec{r}) \equiv \sigma(r)$, and $\sigma(r)$ is given by a set of concentric shells with laterally uniform conductivity values, eqs. (\ref{eq:maxwell_mode_1}-\ref{eq:maxwell_mode_2}) can be solved analytically. The analytical formulae are lengthy and not repeated here, but can be found in \citet{parkinson1983}, among others.

The presented formalism allows us to express any external time-varying magnetic field, that reaches a planet, in terms of an equivalent current density. Therefore, the time series of $q_n^m(t), s_n^m(t)$ coefficients in conjunction with spherical harmonic functions offer a consistent way to impose an arbitrary spatiotemporal external forcing. Every external coefficient represents an independent inducing mode that contributes to the total inducing and induced currents. The latter can be used in eqs. (\ref{eq:b_field_sum}-\ref{eq:e_field_sum}) to obtain a distribution of the EM field throughout the body.  Our physical model implies that the surface of a planet is coupled to external source currents through EM induction. This is a reasonable assumption for planets with electrically neutral atmospheres. If the atmosphere is very thin or fully ionized, galvanic coupling can develop between the interior and exterior, facilitating additional interior-exterior EM energy exchange. The description of such galvanic coupling is generally complicated and will depend on quickly varying properties of ionized material around a planet, which is poorly constrained.  Nevertheless, inductive coupling in such settings will still exist, albeit in the presence of galvanic interactions. 

\section{Magnetic field perturbations due to stellar flares} 
\label{app:perturbations}

The mathematical description (Appendix~\ref{app:induction}) offers a general tool for calculating the electric field distribution within a planet given a time-varying external (inducing) field $\mathbf{B}^{\text{ext}}(\vec{r}, t)$ that is imposed on the planet. We assume that the external field is uniform across the planet and thus can be represented by a linear combination of degree one SH coefficients.  Given that we assume radial symmetry of our conductivity models (no lateral variation due to $\phi$ or $\theta$), it suffices to consider only one coefficient. Without loss of generality, we chose to work with the $P_1^0$ function and corresponding coefficient $q_1^0$.  The assumption of a homogeneous external field is justified by Earth, where the dominant spatial component of the external magnetic field at long periods during global geomagnetic storms is described by a degree one field \citep{parkinson1983}. Nevertheless, higher degree and order spatial harmonics will likely be present in $\mathbf{B}^{\text{ext}}$ owing to the complexity of an ICME field and the interactions with potential magnetospheres and ionospheres \citep{finlay2017challenges}.  However, there are no observational constraints on exoplanets that justify introducing these additional complications into our model.

Having defined the spatial structure of the external magnetic field, we need to prescribe its temporal profile for an ICME, which is described by the time series of the selected SH coefficient, hereinafter referred to as $M(r_p, t)$, where $r_p$ denotes the distance of a planet from the star.  Hence temporal profiles are individual for each planet. Solar system observations and MHD simulations suggest that, to first order, the temporal profile of an ICME magnetic field \citep{verbeke2019evolution} or associated global magnetic storms \citep{temerin2006dst} around planets can be described by Gaussian or exponential decay functions, respectively. Assuming the time origin at $t_0$, the temporal profile of magnetic field perturbations due to an ICME is then calculated as
\begin{equation}
\label{eq:gaussian}
    M(r_p,E_{flare},t) = M_{peak}(r_p,E_{flare})\exp\left(-\frac{(t-t_0)^2}{2\tau^2}\right)
\end{equation}
or
\begin{equation}
\label{eq:exponential}
    M(r_p,E_{flare},t) = M_{peak}(r_p,E_{flare}) H(t-t_0)\exp\left(-\frac{t-t_0}{\tau}\right)
\end{equation}
for Gauss and exponential models, respectively. In both cases, the constant $\tau$ determines the characteristic length of a flare and typically ranges from a few hours to a day. The scaling factor $M_{peak}$ determines the maximum amplitude of the magnetic field perturbation at a given distance $r_p$ and flare energy $E_{flare}$. The time series of $M(r_p,E_{flare},t)$ can be introduced in eqs. (\ref{eq:b_field_td}-\ref{eq:e_field_td}) to calculate the distribution of the EM field within a planet at any given time and location.

\section{Generation and propagation of Coronal Mass Ejections} 
\label{app:gen_and_prop_CME}

\begin{figure}
\centering
\gridline{\fig{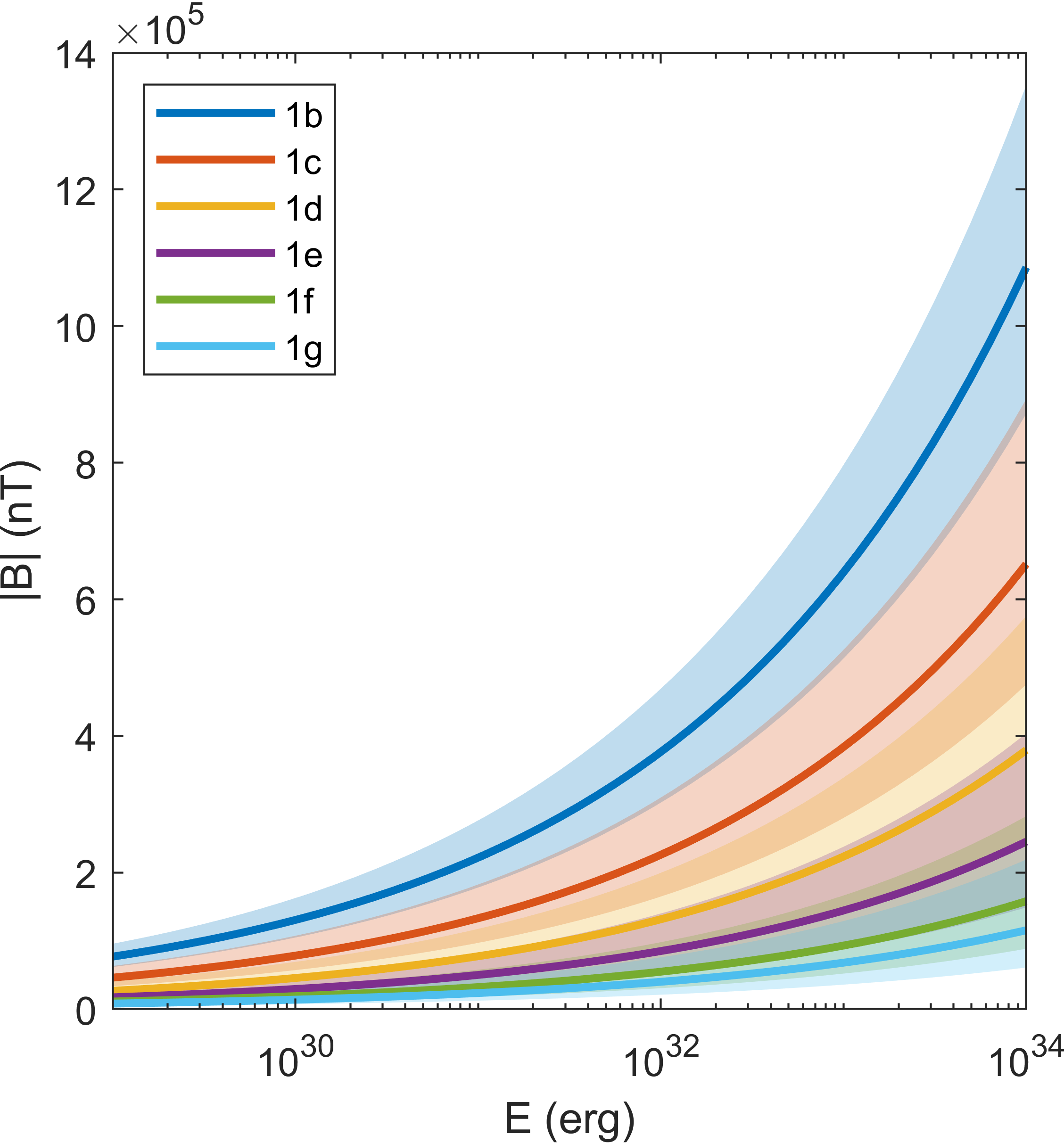}{0.42\textwidth}{(a)}
          \fig{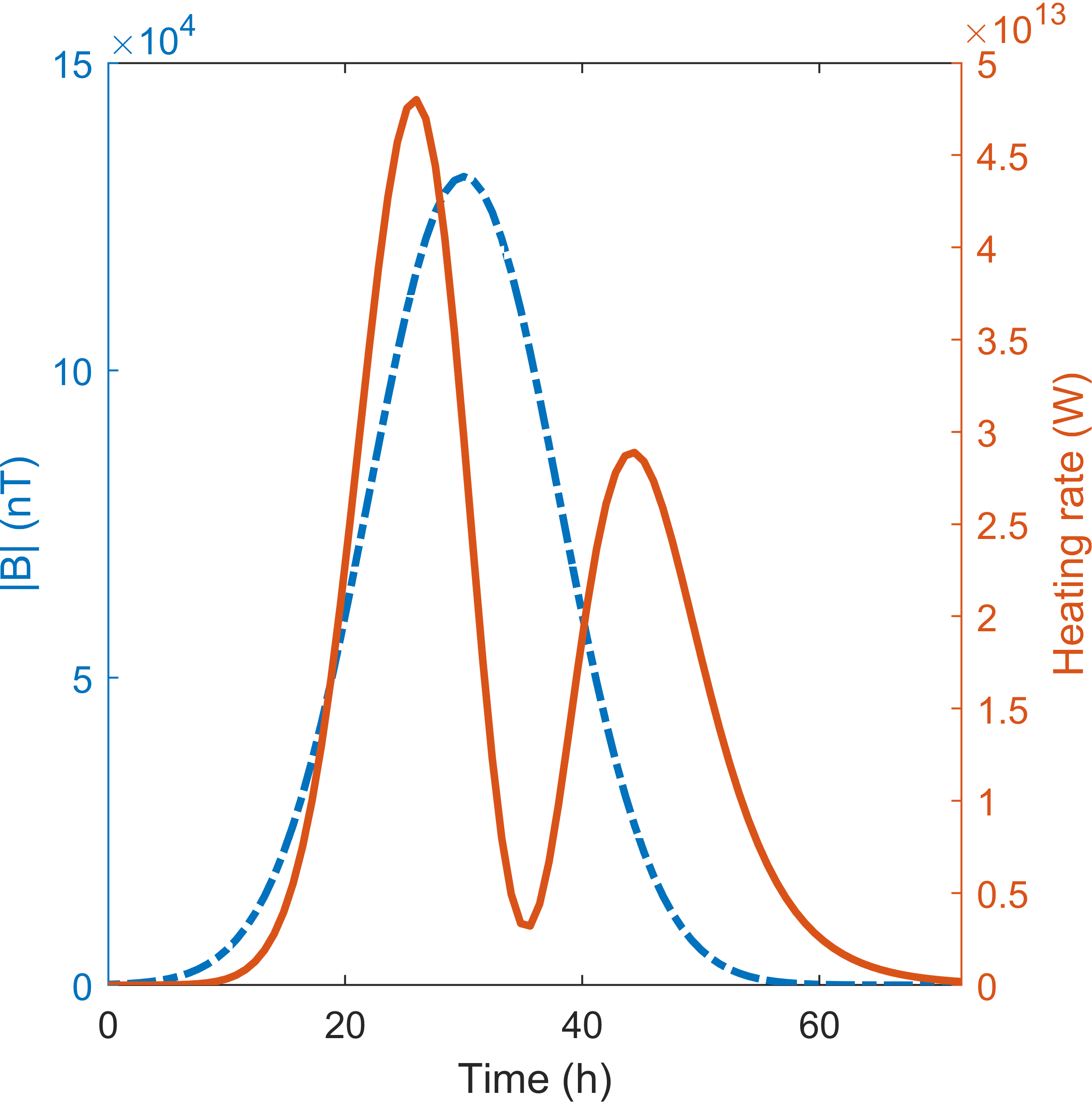}{0.45\textwidth}{(b)}
          }
\caption{(a): Magnetic field strength at each planet as a function of flare energy. The shaded range represents three standard deviations interval in the field decay rate $\gamma \in \mathcal{N}(1.6, 0.1)$ (see Table \ref{tab:params}). (b): Realization of an external magnetic field profile at the TRAPPIST-1d planet for an ICME event with the energy $10^{32}$ erg (dashed line, left axis) and heat dissipated within the TRAPPIST-1d planet assuming a homogeneous 100 $\Omega$m conductivity (solid line, right axis). 
\label{fig:field_vs_heat}}
\end{figure}

In a relevant energy range, stellar flares follow a continuous power law distribution \citep{clauset2009power} with a probability density function (PDF) 
\begin{equation}
    p(E_{flare}) = CE_{flare}^{-\alpha},\label{eq:power_law_pdf}
\end{equation}
where $E_{flare}$ [erg] is the flare (bolometric) energy and normalization constant
\begin{equation}
    C = (\alpha - 1)E_{min}^{\alpha-1}.
\end{equation}
Here, $E_{min}$ is the minimum energy at which the power law still holds. In practice, it is useful to also consider the complementary Cumulative Distribution Function (CDF)
\begin{equation}
    P(E_{flare}) = \left(\frac{E_{flare}}{E_{min}}\right)^{1 - \alpha},
\end{equation}
which defines a probability of an event with an energy $\ge E_{flare}$. A log-log plot of $P(E_{flare})$ versus $E_{flare}$ is commonly used to visualize the flare frequency distribution.  Once $E_{min}$ and $\alpha$ are constrained through observations \citep{paudel2018k2}, we can reproduce the flaring activity of a star by sampling from the distribution (Eq. \ref{eq:power_law_pdf}) over a given period of time. For a flare of a given bolometric energy, the magnetic field strength at a radius $r_p$, measured from the center of a star can be calculated as
\begin{equation}
\label{eq:flare_mag}
    M_{ICME}(E_{flare}, r_p) = M_{\star}(E_{flare})\left(\frac{10R_{\star}}{r_p}\right)^{\gamma},
\end{equation}
where $r_p$ is the distance from the center of a star, $M_{\star}$ [T] is the magnetic field strength at the distance $10R_{\star}$, and $\gamma$ is the magnetic field decay rate. The model in Eq. (\ref{eq:flare_mag}) is adopted from \cite{samara2021readily}, and factor $M_{\star}$, representing the magnetic field strength in the near-star zone, is derived from their mean model \citep[Appendix A,][]{samara2021readily}. Following the analysis of extensive observational and MHD simulation data \citep{samara2021readily}, the decay rate $\gamma$ was estimated to lie in the range $[1.2, 2.0]$, with the most probable value of $1.6$ for our solar system. We have verified this model independently by comparing it with the geoeffective solar CME events from the LASCO catalogue \citep{gopalswamy2009soho} and found a good agreement. Figure \ref{fig:field_vs_heat}a shows range of magnetic field strength at planets in this study. 

Not every ICME will collide with a planet. To calculate the frequency of impact, we adopt the model of \citep{khodachenko2007coronal}:
\begin{equation}
\label{eq:prob_impact}
f_{Impact}(\Delta_{CME}) = \frac{\left(\Delta_{CME} + \delta_{pl}\right)\sin{\left[\left(\Delta_{CME}+\delta_{pl}\right)/2\right]}}{2\pi\sin\Theta}
\end{equation}
where $\Delta_{CME}$ is the angular size of a CME, $\Theta = 80^{\circ}$ is the range of active stellar latitudes where CMEs occur. The solid angle subtended by a planet $\delta_{pl}$ plays virtually no role due to its small value ($ \lessapprox 10^{-5}$), but was included for completeness. Since flare-associated ICMEs have a finite extent in space and propagate at a finite speed, studies that ignore the frequency of an impact implicitly assume that every ICME impacts a planet, implying that flares must occur so often that no matter where a planet is, it will always be within an ICME. However, none of the observed TRAPPIST-1 flares satisfy this `persistent impact' condition. Therefore, we use Eq. (\ref{eq:prob_impact}) to calculate the collision probability for the TRAPPIST-1 system. We do not fix the value of $\Delta_{CME}$, but rather sample it over a plausible range of values (Table \ref{tab:params}). Taking the mean value of $\bar{\Delta}_{CME} = 60^{\circ}$, we obtain the average frequency of impact $\bar{f}_{Impact} = 0.084$.

\begin{deluxetable}{ccl}
\tablecaption{Prior probability distributions of randomly sampled parameters.\label{tab:params}}
\tablewidth{0pt}
\tablehead{
\colhead{Parameter} & \colhead{Distribution} & \colhead{Description}
}
\decimalcolnumbers
\startdata
$\Delta_{CME}$ & $\mathcal{U}(40,80)$ & Angular size of a CME (deg) (Appendix \ref{app:gen_and_prop_CME}) \\
$\bar{N}_f$ & $\mathcal{U}(141,262)$ & Number of ICME events for the TRAPPIST-1 star per year (Section \ref{sec:methods}) \\
$\tau$ & $\mathcal{U}(2,24)$ & ICME duration coefficient (hr) (Appendix \ref{app:perturbations}) \\
$\gamma$ & $\mathcal{N}(1.6,0.1)$ & Decay rate of the ICME magnetic field strength (Appendix \ref{app:gen_and_prop_CME}) \\
\enddata
\tablecomments{$\mathcal{U}$ and $\mathcal{N}$ denote uniform and normal distributions, respectively. The arguments specify sampled ranges or mean and standard deviation values for uniform and normal distributions, respectively.}
\end{deluxetable}

\section{Induction heating}
\label{app:heating}

Multiplying equations (\ref{eq:maxwell_time1}) and (\ref{eq:maxwell_time2}) by $\mathbf{E}$ and $\mathbf{H}$, respectively, and subtracting yields
\begin{equation}
\left( \nabla\times \mathbf{H} \right)\cdot \mathbf{E} - \left( \nabla\times\mathbf{E} \right)\cdot \mathbf{H} = \mathbf{J}\cdot\mathbf{E} + \left(\frac{\partial\mathbf{B}}{\partial t}\right)\cdot \mathbf{H}.
\end{equation}
Using $\left( \nabla\times \mathbf{H} \right)\cdot \mathbf{E} - \left( \nabla\times\mathbf{E} \right)\cdot \mathbf{H} = -\nabla\cdot\left( \mathbf{E}\times\mathbf{H} \right)$, integrating over volume and denoting the last term as $\frac{\partial W}{\partial t}$, we obtain the Poynting theorem
\begin{equation}
\label{eq:poynting}
    -\iiint_V \frac{\partial W}{\partial t}\textnormal{d}V = \oiint_{\partial V} \left(\mathbf{E}\times\mathbf{H}\right)\cdot \mathbf{\hat{n}} \textnormal{d}S + \iiint_V \mathbf{E}\cdot\mathbf{J} \textnormal{d}V,
\end{equation}
which expresses the rate of change of electromagnetic energy in a volume $V$. The first term on the right-hand side defines the energy radiated away through the surface of a planet. The second term represents the amount of energy dissipated within the volume (e.g. due to Joule heating). Physical variables in Eq. (\ref{eq:poynting}) are time-varying functions of space. Therefore, we can express the time-averaged volumetric heat energy as
\begin{equation}
\label{eq:diss_power}
    Q_{avg} = \frac{1}{t_2 - t_1}\int_{t_1}^{t_2}\int_{V} \left[\mathbf{E} \cdot \mathbf{J}\right] \textnormal{d}V\textnormal{d}t,
\end{equation}
where integrated quantities are functions of space and time. 

It is instructive to consider a simple scenario of a homogeneous sphere of radius $R$ and conductivity $\sigma$. Assuming that currents within the sphere are induced by a homogeneous harmonic external field (i.e., $n = 1$) of amplitude $M_{peak}$, we can express the dissipated power at angular frequency $\omega$ as \citep{bromley2019ohmic}
\begin{eqnarray}
\label{eq:diss_power_sphere}
  Q &=& \frac{1}{2\sigma}\int_V \left| \mathbf{J}\right|^2 \textnormal{d}V \nonumber \\
  &=& \frac{\pi R^3 \omega M_{peak}^2}{\mu} \textnormal{Im}\left[ \frac{j_2(kR)}{j_0(kR)} \right],
\end{eqnarray}
where $j_n(\eta)$ is the spherical Bessel function of the first kind, $k = \sqrt{2\textnormal{i}}/\delta_s$ is wavenumber and $\delta_s = \sqrt{2/\omega\mu\sigma}$ the skin depth. For a case when $R/\delta_s \leq 1$, we can user Taylor expansion of the ratio of Bessel functions around zero and approximate Eq. \ref{eq:diss_power_sphere} as 
\begin{eqnarray}
\label{eq:diss_power_sphere_approx}
    Q &\approx& \frac{\pi R^3 \omega M_{peak}^2}{\mu} \textnormal{Im}\left[ \frac{(kR)^2}{15} + \frac{(kR)^6}{1575} \right] \nonumber \\
    &\approx& \frac{\pi\sigma R^5 \omega^2 M_{peak}^2}{15} \left[1 - \frac{4R^4}{105\delta_s^4} \right].
\end{eqnarray}

Eqs. \ref{eq:diss_power_sphere}-\ref{eq:diss_power_sphere_approx} above are for a single angular frequency $\omega$. The actual temporal profiles of inducing currents due to CMEs are transient signals approximated by Eqs. \ref{eq:gaussian}-\ref{eq:exponential}. Therefore, to calculate the dissipated power for a transient signal, the integration (summation) over a finite time period (respectively, spectrum) is performed in Eq. \ref{eq:diss_power}. An example of a dissipated energy profile for a transient CME event is shown in Figure \ref{fig:field_vs_heat}b.

\section{Interior temperature}
\label{app:boundarylayer}
We invoke boundary layer theory (BLT) \citep[e.g.,][]{TS14} to derive the interior temperature of the TRAPPIST-1 planets from the mean flare-induced heat power density (e.g., layered model in Figure~\ref{fig:profiles}) in addition to radiogenic heat.  BLT provides an approximate solution to the fluid equations when the interior is vigorously convecting.  For a planet dominantly heated within, the Rayleigh number of the mantle is

\begin{equation}
    {\rm Ra}_H=\frac{a \rho g H d^5}{k \nu \kappa},
\end{equation}
where $a$ is the thermal expansion, $\rho$ is the density, $g$ is the gravitational acceleration, $H$ is the internal heating rate per unit mass, $d$ is the thickness of the mantle, $k$ is the thermal heat conductivity, $\nu$ is the kinematic viscosity, and $\kappa$ is the thermal diffusivity.  The following parameters are fixed to Earth-like values: $a = 3 \times 10^{-5}$ K$^{-1}$, $k=4$ Wm$^{-1}$K$^{-1}$, $\nu \rho=10^{21}$ Pa$\cdot$s, and $\kappa=10^{-6}$ m$^2$s$^{-1}$.  For each TRAPPIST-1 planet, $d$ is $0.52$ of the planetary radius \citep{agol2021refining}.  The total heating rate $H$ can be a combination of an assumed Earth-like radiogenic component ($H_r=9 \times 10^{-12}$ Wkg$^{-1}$) and flare-induced ($H_f$) heating.  Flare-induced heating is determined by integrating the heat power density within the mantle and normalising by the mantle mass.  The PREM is assumed to give the density profile for each planet from which an average mantle density ($\rho$) is calculated and the surface gravity ($g$) is determined from observations that constrain the planetary mass and radius \citep{agol2021refining}.  BLT provides a relationship between ${\rm Ra}_H$ and the temperature drop from the interior ($T_1$) to the surface ($T_0$):
\begin{equation}
    T_1-T_0 = C \frac{\rho H d^2}{k} {\rm Ra}_H^{-1/4},
    \label{eq:tempdrop}
\end{equation}
where $C$ is a prefactor that can be derived from BLT and is 1.49 for a transient boundary layer and 2.45 for a steady-state boundary layer.  The larger estimate of $C=2.45$ recovers Earth's interior--surface temperature drop of around 2000 K to within a factor of two \citep{TS14}.  Here, we are preoccupied with estimating the additional temperature increase of the mantle ($\Delta T$) due to both flare and radiogenic heating compared to only radiogenic heating:
\begin{equation}
    \Delta T= C \frac{\rho d^2}{k} \left(H_{r+f}{\rm Ra}_{H_{r+f}}^{-1/4} - H_{r}{\rm Ra}_{H_{r}}^{-1/4} \right).
\end{equation}
The input parameters and results are summarised in Table~\ref{tab:boundarylayer} and discussed in the main text in Section~\ref{sec:discussion}.

\movetabledown=20mm
\begin{deluxetable*}{l|lll|l|l}
\label{tab:boundarylayer}
\tablecaption{Increase in the interior temperature of the TRAPPIST planets due to flare heating and radiogenic heating compared to only radiogenic heating ($\Delta T$).}
\tablehead{
\multicolumn{1}{c|}{Planet} & \colhead{$\rho$ (mantle)} & \colhead{$g$} & \colhead{$d$} & \multicolumn{1}{|c|}{No magnetic field} & \multicolumn{1}{c}{With magnetic field}\\
& (kgm$^{-3}$) & (ms$^{-2}$) & (km) & $\Delta T$ (K) & $\Delta T$ (K)}
\startdata
b & 4622 & 10.8 & 3720 & 62 & 1875\\
c & 4606 & 10.65 & 3655 & 23  & 759\\
d & 4373 & 6.11 & 2900 & 10 & 341\\
e & 4455 & 8.01 & 3070 & 4 & 133\\
f & 4563 & 9.32 & 3485 & 1 & 54\\
g & 4633 & 10.15 & 3765 & 1 & 28
\enddata
\tablecomments{'No magnetic field' and 'With magnetic field' refer to an Earth-like intrinsic planetary magnetic field. The mean estimates of the produced heat were used.}
\end{deluxetable*}


\end{document}